\documentclass[11pt]{article}

\usepackage{amsmath}
\usepackage{amssymb}
\usepackage{amsthm}
\usepackage{bbm}
\usepackage[margin=1in]{geometry}
\usepackage{authblk}
\usepackage{url}
\usepackage{hyperref}
\usepackage{natbib}
\usepackage{graphicx}
\usepackage{booktabs}
\usepackage{enumitem}
\usepackage{algorithm}
\usepackage{algpseudocode}

\usepackage{secdot}

\usepackage[font=footnotesize,labelfont=bf,labelsep=period]{caption}

\newtheorem{theorem}{Theorem}
\newtheorem{proposition}[theorem]{Proposition}

\newtheorem{corollary}[theorem]{Corollary}

\theoremstyle{definition}

\newtheorem*{remark}{Remark}

\newcommand{\CI}{\text{CI}}
\newcommand{\ind}{\perp\!\!\!\!\perp} 
\newcommand{\E}{\textit{e}}

\newcommand{\blind}{1}
 
\begin{document}
	
	\title{Anytime-Valid Tests of Conditional Independence Under Model-X}
	\author[12]{Peter Gr\"unwald}
	\author[3]{Alexander Henzi}
	\author[12]{Tyron Lardy}
	\affil[1]{Centrum Wiskunde \& Informatica, Amsterdam, The Netherlands}
	\affil[2]{Leiden University, Leiden, The Netherlands}
	\affil[3]{ETH Z\"urich, Switzerland}
	\affil[ ]{{\tt pdg@cwi.nl}, {\tt alexander.henzi@stat.math.ethz.ch}, {\tt t.d.lardy@math.leidenuniv.nl}}
	\date{\today} 
	\maketitle
	
	
	\begin{abstract}
        We propose a sequential, anytime-valid method to test the conditional independence of a response $Y$ and a predictor $X$ given a random vector $Z$.
        The proposed test is based on \E-statistics and test martingales, which generalize likelihood ratios and allow valid inference at arbitrary stopping times. 
        In accordance with the recently introduced model-X setting, our test depends on the availability of the conditional distribution of $X$ given $Z$, or at least a sufficiently sharp 
        approximation thereof.
        Within this setting, we derive a general method for constructing \E-statistics for testing conditional independence, show that it leads to growth-rate optimal \E-statistics for simple alternatives, and prove that our method yields tests with asymptotic power one in the special case of a logistic regression model. 
        A simulation study is done to demonstrate that the approach is competitive in terms of power when compared to established sequential and nonsequential testing methods, and robust with respect to violations of the model-X assumption.
		
	\end{abstract}
	
	\section{Introduction} \label{sec:intro}

	A fundamental task in many areas of research, such as economics, genetics, and pharmacology, is to find out whether there is an association between a response $Y$ and an explanatory variable $X$, given a vector of covariates $Z$.
	Mathematically, 
	absence of
	such an association is defined as conditional independence (CI) between $X$ and $Y$ given $Z$, denoted by $X\ind Y\mid Z$ \citep{Dawid1979}.
	A standard way to tackle these problems is to assume a (semi-)parametric model on $Y$ given $X$ and $Z$, encoding the dependence between $Y$ and $X$ in a model parameter.
	For example, in the logistic model the probability that a binary random variable $Y$ equals one is regressed on $(X,Z)$, and the regression coefficient corresponding to $X$ is zero if and only if $X\ind Y\mid Z$. 
	Within these parametric models, there are well established methods to test conditional independence, such as the likelihood ratio test for generalized linear models~\citep{mccullagh1989generalized}. 
	However, all of these tests  have in common that they fail to uphold a type-I error guarantee when the model assumptions are not satisfied.
	If $Z$ is continuously distributed such error inflation is in fact unavoidable: unless further assumptions on the distribution of $(X,Y,Z)$ are imposed, there exist no nontrivial tests of CI which maintain type-I error guarantees \citep{Shah_2020}.
	
	However, \citet{Candes2018} show that given additional knowledge, i.e.~the distribution of $X$ conditional on $Z$, nontrivial tests of conditional independence can be designed without further assumptions on the distribution of $(X,Y,Z)$.
	This has been dubbed the Model-X (MX) setting, and the condition that the distribution of $X$ conditional on $Z$ is known is called the Model-X (MX) assumption, though there are settings where the MX `assumption' is \emph{known} to be true.
	Perhaps the most prominent example is a randomized clinical trial, where the distribution of treatment/control is imposed by the researchers. 
	Another example is conjoint analysis, which is a survey-based experiment where respondents are asked to express a preference between multiple hypothetical products with different attributes.
	These attributes are randomized according to a distribution chosen by the researchers, with the aim to find out whether one or more of the attributes have an influence on consumer preference (see e.g.~\cite{ham2022using}). 
	Furthermore, \citet{Candes2018} show that there are ample scenarios where at least an accurate estimate of the conditional distribution of $X$ given $Z$ is known, while also acknowledging that the MX assumption might not be appropriate if this is not the case.
	
	Under the MX assumption, \citeauthor{Candes2018} derive the conditional randomization test (CRT), which  has nontrivial power to detect CI.
	Recently, much effort has gone into relaxing the MX assumption and improving the robustness of the CRT under misspecification of the conditional distribution of $X\mid Z$~\citep{Katsevich2020OnModel-X,Li2022MaxwayCRT,niu2022reconciling}.
	The CRT and most of its extensions have in common that they are based on p-values computed on batches of data, and therefore designed for fixed sample size experiments.
	In this article, we focus on anytime-valid tests for CI under MX. 
	Anytime-valid tests allow for more flexibility when testing compared to the nonsequential CRT
    and also allow covariate adaptive designs, where the distribution of $X$ does not only depend on $Z$, but also on past data, such as in response-adaptive sampling schemes.
	Previous work on anytime-valid tests of CI under MX has been done by \citet{Duan2022}, who propose tests for the case that $X$ is binary.
	Their approach is based on pragmatic game-theoretic principles, whereas in this article, we aim to describe and theoretically analyze anytime-valid tests for CI for general $X$. \citet{Shaer2022} have worked on an extension of the work by \citet{Duan2022} concurrently, and we will discuss their work and the connections to this article in Section \ref{sec:discuss}.
	
	Our hypothesis tests are based on the concept of \E-statistics\footnote{E-statistics are most commonly known as \E-variables; we refer to them as \E-statistics to stress the data dependence.}~\citep{Gruenwald2019,RamdasRLK21,Vovk2020E-values:Applications}. 
    \textit{E}-statistics have been introduced as an alternative to p-values that is inherently more suitable for testing under optional stopping and continuation \citep{Gruenwald2019,Vovk2020E-values:Applications,Ramdas2020}. While they have their roots in  the work on anytime-valid testing by H. Robbins and students (e.g. \citep{darling1967confidence}), interest in them has exploded in recent years; see, for example, also 
    \citet{Shafer2021,Gruenwald2022_neymanpearson}.
    \textit{E}-statistics can be associated with a natural notion of optimality, called GRO (growth-rate optimality) by \cite{Gruenwald2019}, which may be viewed as an analogue of statistical power in an optional stopping context. 
	We derive a general method for constructing \E-statistics for conditional independence testing under model-X and
    show that the method that we propose is optimal in this GRO sense for testing conditional independence against point alternatives under MX.
	This result should be seen as an anytime-valid analogue to the result by~\cite{Katsevich2020OnModel-X}, who use the Neyman-Pearson lemma to derive test statistics for which the conditional randomization test is the most powerful conditionally valid MX CI test.
	Furthermore, we show that under misspecification of the distribution of $X$ given $Z$, our method retains type-I error sequentially just as well as the CRT does for blocks of data~\citep{Berrett2020}. 
	Finally, we discuss in detail an application to the setting where $Y$ is binary, where we use logistic regression to construct an anytime-valid test of conditional independence. 
	Under the MX assumption, this test is valid even if the logistic model assumptions are not satisfied, and when they are, it is guaranteed to have asymptotic power one. 
	
	The rest of this article is structured as follows. In Section~\ref{sec:background} we discuss the background needed for this paper. 
	This includes an introduction to the conditional randomization test in Section~\ref{sec:crt} and to \E-statistics in Section~\ref{sec:e_stat}. 
	In Section~\ref{sec:e_crt} we discuss in detail our anytime-valid method of testing conditional independence under MX. 
	This includes an analysis of optimality in Section~\ref{sec:optimality}, a bound on the worst-case rejection rate in Section~\ref{sec:rej_bound}, and an application to binary response variable in Section~\ref{sec:logistic}.
	We compare our method to established tests of conditional independence in a simulation study in Section~\ref{sec:sim}.
	This includes a comparison in terms of type-I error (Sections~\ref{sec:sim_null} and~\ref{sec:robustness}) and in terms of power (Section~\ref{sec:sim_alt}).
	Our method is further highlighted by an application to a real-world data set in Section~\ref{sec:data}.
	This article is concluded with a discussion of our results in Section~\ref{sec:discuss}.
	All proofs are deferred to Appendix~\ref{app:proofs}.
	
	\section{Background}\label{sec:background}
	In this section, we give a brief introduction to the conditional randomization test, as well as to \E-statistics.
	Throughout this section, as well as the rest of the article, we assume that the data consists of independent and identically distributed (i.i.d.) tuples $D_i=(X_i,Y_i,Z_i)\in \mathcal{D}:=\mathcal{X}\times\mathcal{Y}\times\mathcal{Z}$.
	In the section on the CRT, we assume that data comes in as a single block $D^n = (D_1, \dots, D_n)$ of $n\in \mathbb{N}$ data points, so that $i$ above ranges from $1$ to $n$.
	In contrast, in the section on \E-statistics, we assume that the data comes in sequentially as a stream $(D_n)_{n\in \mathbb{N}}$. We use the notation $f_{Y\mid X,Z}(y\mid x,z)$ to denote the conditional density of $Y$ given $X$ and $Z$ evaluated at $(x,y,z)$ when the joint density of $(X,Y,Z)$ is $f$, and analogous notation for other conditional densities, e.g.~$f_{X\mid Z}(x\mid z)$ for the density of $X$ given $Z$.
	
	\subsection{The Conditional Randomization Test}\label{sec:crt}
	The conditional randomization test by~\citet{Candes2018} works under the MX assumption, i.e.~that the distribution of $X\mid Z$ is known.
	This holds, for example, when $X$ corresponds to a randomized treatment in a clinical trial, since the researchers specify the randomization mechanism themselves. 
	\citet{Candes2018} give other examples where the this distribution is at least known to a much higher precision than the joint distribution of $(X, Y, Z)$ because much more samples of pairs $(X, Z)$ are available than data including $Y$. 
	We let $Q_z$ denote the distribution of $X$ given that $Z=z$.
	We are interested in testing $X\ind Y \mid Z$, which in the MX setting corresponds to the null hypothesis
	\begin{equation}\label{eq:null_hypothesis}
		\mathcal{H}_0=\left\{P\in \mathcal{P}(\mathcal{D}): X\overset{P}{\ind}Y\mid Z, \ P_{X|Z} = Q_Z\right\},
	\end{equation}
	where $\mathcal{P}(\mathcal{D})$ denotes the set of all probability distributions on $\mathcal{D}$, and $P_{X|Z}$ is a shorthand for the conditional law of $X$ given $Z$.
	The starting point of the CRT is to choose any test statistic $T: \mathcal{D}^n \mapsto \mathbb{R}$ that measures the dependence of $X$ and $Y$ given $Z$, so that larger values of $T$ indicate stronger dependence. 
	Since the conditional distribution of $X$ given $Z$ is known, one can simulate independent new realizations $\tilde{X}_{j,i} \sim Q_{Z_i}$, $j = 1, \dots, M$, $i = 1, \dots, n$, so that under the null hypothesis $\tilde{X}^{n}_j = (X_{j,1}, \dots, X_{j,n})$ and $X^n$ have the same distribution conditional on $Y^n$ and $Z^n$.
	Hence the triplets $(\tilde{X}^{n}_j, Y^n, Z^n)$, $j = 1, \dots, M$, and $(X^n, Y^n, Z^n)$ are exchangeable, and
	\begin{equation}\label{eq:mcpvalue}
		p_M(X^n, Y^n, Z^n) = \frac{1 + \sum_{j=1}^M 1\{T(\tilde{X}^{n}_j, Y^n, Z^n) \geq T(X^n, Y^n, Z^n)\}}{1+M}
	\end{equation}
	is a ``Monte Carlo'' p-value for the null hypothesis \eqref{eq:null_hypothesis}. 
	More precisely, for any distribution $P\in\mathcal{H}_0$,
	\[
	P(p_M(X^n, Y^n, Z^n) \leq \alpha) \leq \alpha.
	\]
	In the case that the distributions $Q_z$ are not known exactly, but only some estimate $\hat{Q}_z$, an obvious question is how robust a test based on $\hat{Q}_z$ may be. 
	\citet[Theorem 4]{Berrett2020} show that 
	\begin{equation} \label{eq:crt_robustness}
		P(p_M(X^n, Y^n, Z^n) \leq \alpha \mid Y^n, Z^n) \leq \alpha + d_{\mathrm{TV}}(\hat{Q}^{n}_{Z^n}, Q^{n}_{Z^n}),
	\end{equation}
	where $Q^{n}_{Z^n}$ is the product of the distributions $Q_{Z_i}$, $i = 1, \dots, n$, and $d_{\mathrm{TV}}$ the total variation distance. Precise estimation of conditional distributions in total variation distance is admittedly a challenging problem, but also not completely unrealistic, as discussed among others by \citet[Section 5.1]{Berrett2020}.
	We furthermore give a brief discussion of available literature on estimation in terms of KL divergence at the end of Section~\ref{sec:optimality}, and a bound on KL gives a bound on total variation.
	
    There remains the question how to choose the statistic $T$.
    \citet{Katsevich2020OnModel-X} show that for a point alternative with density $f$ for $(X,Y,Z)$, using the conditional density $f_{Y\mid X,Z}$ of $Y$ given $(X,Z)$ as statistic $T$ leads to the most powerful conditionally valid test against $\mathcal{H}_0$.
	This result suggests that a reasonable method is to use an estimator $\hat f_{Y\mid X,Z}$ of the true conditional density as statistic.
	Importantly, any estimation error does not lead to a loss of type-I error guarantee, but only to a decrease in power, as the p-value in~\eqref{eq:mcpvalue} is valid for \emph{any} statistic. 
	Alternatively, one could use any measure of conditional independence, e.g.~the absolute value of the fitted coefficient in a lasso regression model, as originally proposed by~\citet{Candes2018}. 
	A downside to this method is that it is computationally expensive, as it requires a lasso model to be fit for the original data as well as for all the simulated data points. 
	\citet{liu2021fast} propose a ``leave-one-covariate-out'' variant of these lasso statistics, which is significantly less computationally demanding, while leading to similar power.
	
	\subsection{\textit{E}-statistics, Test Martingales and Anytime-Valid Tests}\label{sec:e_stat}
	An \E-statistic is any function of the data $S_n:\mathcal{D}^n\rightarrow [0,\infty)$ such that $\mathbb{E}_P[S_n(D^n)] \leq 1$ for all $P\in\mathcal{H}_0$. 
	An \E-statistic evaluated on a realization of the data will be referred to as an \E-value.
	Previously, \E-statistics and \E-values have appeared in the literature as likelihood ratios (although the concept is vastly more general than such ratios), particular Bayes factors and betting scores~\citep{Shafer2021}.
	Large \E-values constitute evidence against the null hypothesis, since $P(S_n(D^n) \geq 1/\alpha) \leq \alpha$ by Markov's inequality, so that the type-I error of the test $1\{S_n(D^n)\geq 1/\alpha\}$ is bounded by $\alpha$.
	However, such a test is defined for a block of data $D^n$. 
	In a sequential setting, one instead observes a stream of data $(D_n)_{n\in \mathbb{N}}$. 
	We therefore define, more generally, a sequence of conditional \E-statistics as a sequence of statistics $(E_n(D^n))_{n\in \mathbb{N}}$, such that
	\[
	\mathbb{E}_P[E_n(D^n)\mid D^{n-1}]\leq 1
	\]
	for all $n\in\mathbb{N}$ and $P\in\mathcal{H}_0$.
	For $n=1$, the expectation is supposed to be read unconditionally. 
	Intuitively, the conditional \E-statistic at time $n$ measures the evidence in round $n$ against $\mathcal{H}_0$ conditional on the past data, and the cumulative product of these conditional \E-statistics $S_n(D^n)=\prod_{i=1}^n E_i(D^i)$ is a measure of the total accumulated evidence against the null hypothesis.
	Formally, the sequence $(S_n(D^n))_{n\in\mathbb{N}}$ of cumulative products forms a nonnegative supermartingale with starting value bounded by $1$, a so-called test martingale, i.e.
	\begin{equation}
		\mathbb{E}_P[S_{n+1}(D^{n+1}) \mid D^n] \leq S_{n}(D^{n})
	\end{equation}
	for all $P \in \mathcal{H}_0$ and $n \in \mathbb{N}$, and $\mathbb{E}_P[S_1(D_1)]\leq 1$. 
	By Ville's inequality \citep[see e.g.][]{Shafer2021}, such test martingales satisfy for any $\alpha>0$
	\begin{equation} \label{eq:ville}
		P(\exists \, n \in \mathbb{N}\colon S_n(D^n) \geq 1/\alpha) \leq \alpha.
	\end{equation}
	A sequential test can thus be defined by monitoring $S_n(D^n)$ and rejecting if it exceeds $1/\alpha$. The inequality~\eqref{eq:ville} ensures that this test retains type-I error control, no matter how we choose the moments to peek at $S_n(D^n)$. 
    In fact, \eqref{eq:ville} is easily derived from a more basic property: 
    $(S_n(D^n))_{n \in \mathbb{N}}$ satisfies $\mathbb{E}_P[S_{\tau}(D^\tau)] \leq 1$ for any stopping time $\tau$, that is, the stopped process $S_{\tau}(D^\tau)$ is again an \E-statistic, both for data dependent and externally imposed stopping rules~\citep[Proposition~2]{Gruenwald2019}. 
    Tests with this property will be referred to as anytime-valid tests.
	
	Perhaps the most prominent example of a test martingale is the likelihood ratio process $L_n = \prod_{i=1}^n p_1(Y_i)/p_0(Y_i)$ for testing the null hypothesis that independent $(Y_n)_{n\in \mathbb{N}}$ stem from a distribution with density $p_0$ against the alternative density $p_1$. Tests based on likelihood ratio processes $(L_n)_{n\in \mathbb{N}}$ are called sequential probability ratio test (SPRT) and have first been studied by \citet{Wald1947},
	though Wald, like most of the subsequent sequential analysis literature, uses them with  reject/accept rules different from ours (e.g. one rejects if the value exceeds $(1-\beta)/\alpha$ instead of $1/\alpha$) that preclude optional stopping and require specific stopping times/rules. 
	We refer to \citet[Section 7]{Gruenwald2019} for further discussion of related literature on sequential testing.
	
	Under violations of the null hypothesis, one would hope that an \E-statistic or test martingale attains high values, which gives evidence to reject the null hypothesis. This requires a suitable analogue of power, or notion of optimality, for \E-statistics. We follow \citet{Gruenwald2019,Shafer2021} and try to find \E-statistics that maximize the logarithmic expected value under the alternative.  
	That is, suppose the alternative hypothesis is given by a single distribution $\mathcal{H}_1=\{P^*\}$, then the growth-rate optimal (GRO) \E-statistic is defined as the \E-statistic that maximizes 
	\begin{equation}\label{eq:GRO}
		S_n \mapsto \mathbb{E}_{P^*}[\log S_n(D^n)]
	\end{equation}
	over all \E-statistics. At first glance, this notion of optimality seems counterintuitive for sequential tests, since it is defined for individual \E-statistics and not test martingales. Really, one would hope to find a test martingale $(S_n)_{n\in {\mathbb N}}$  such that $S_{\tau}$ maximizes  (\ref{eq:GRO}) with $n$ replaced by $\tau$, simultaneously for all stopping times $\tau$. 
	Remarkably, in the special setting of this paper, the sequence of GRO statistics $S_1, S_2, \ldots$ that we will consider for a point alternative does have exactly this property, as follows  from Theorem~12 of \citet{koolen2022logoptimal}, providing additional justification for our focus on GRO.
	This is discussed briefly in Appendix~\ref{app:loavev}.
	
	\section{Conditional Independence Testing With \textit{E}-statistics} \label{sec:e_crt}
	The conditional randomization test and its permutation version by \citet{Berrett2020} are defined for batches of data $D^n$. 
	We show here how to create \E-statistics for sequential testing with a stream of data $(D_n)_{n\in\mathbb{N}}$  under the MX assumption.
	Our method can be seen as a broad generalization of an  \E-statistic introduced in the proof of the main theorem of~\citet{Turner2021}, who essentially handle the case that $Y$ and $X$ are both Bernoulli.
	
	\begin{theorem}\label{thm:e_var}
	Let $h_n: \mathcal{D} \to [0,\infty)$, $n \in \mathbb{N}$, be nonnegative measurable functions such that $h_n$ is determined after seeing data $D^{n-1}$. Then the sequence $(E_{h_n}^{\CI}(D_n))_{n\in\mathbb{N}}$ defined by
	\begin{equation}\label{eq:e_randomization}
        E_{h_n}^{\CI} (D_n)=\frac{h_n(X_n,Y_n,Z_n)}{\int_\mathcal{X} h_n(x,Y_n,Z_n) \, \mathrm{d}Q_{Z_n}(x)}.
	\end{equation}
	is a sequence of conditional \E-statistics for the null hypothesis \eqref{eq:null_hypothesis}. Consequently,
		\begin{equation}\label{eq:test_martingale}
		S_{h^n}^{\CI}(D^n)=\prod_{i=1}^n E_{h_i}^{\CI} (D_i)
	\end{equation}
	is a test martingale.
	\end{theorem}
	
    To be explicit, the general workflow of our method is as follows.
    At each time $n = 1, 2, \dots$, a test function $h_n$ is chosen, usually depending on the past data $D^{n-1}$. 
    After seeing the data point $D_n$, the conditional \E-statistic \eqref{eq:e_randomization} is computed and the cumulative product is updated according to $S_{h^n}^{\CI}(D^n) = S_{h^{n-1}}^{\CI}(D^{n-1})\cdot E_{h_n}^{\CI}(D_n)$. 
    For a test at level $\alpha$, one could stop and reject the null hypothesis as soon as $S_{h^n}^{\CI}(D^n) \geq 1/\alpha$ for the first time.

	\begin{remark}
		At this point, it should be noted that the MX assumption is stronger  
		than needed within our sequential setup:
		the MX assumption requires that the data points $(X_n,Y_n,Z_n)$ are i.i.d.~and that the distribution of $X_n$ given $Z_n$ is given by $Q_{Z_n}$.
		However, in our sequential setting it would actually be allowed for the distribution of $(X_n, Y_n, Z_n)$ to depend on the data $D^{n-1}$.
		At each time $n$, we would use in the denominator a distribution $Q_{Z_n, D^{n-1}}$.
		It should be clear that the resulting sequence of random variables still defines a test martingale, but with respect to the altered null hypothesis that $Y_n \ind X_n \mid Z_n, D^{n-1}$ and $X_n \mid Z_n \sim Q_{Z_n, D^{n-1}}$ for all $n \in \mathbb{N}$.
		Clearly, the assumption that $Q_{Z_n, D^{n-1}}$ is known (or can be estimated with high precision) is not realistic in many settings with temporal dependence.
		One example where this extension would be useful are clinical trials with so-called covariate- or response-adaptive designs, where the allocation of patients to a treatment depends either on the imbalance of treatment/control in certain covariate groups or on previously observed responses from patients~\citep{Robbins1952SomeExperiments,Pocock1975SequentialTrial,Zhang2007}.
		To avoid cluttering notation, we will not explicitly denote this potential past data dependence, but it is good to keep in mind. 
	\end{remark}
	
	In most cases in practice, one has to resort to simulations to approximate the integral in the denominator of \eqref{eq:e_randomization}. 
	That is, one simulates $M$ independent $\tilde{X}_1, \dots, \tilde{X}_M$ according to $Q_{Z_n}$ and replaces the integral by the empirical average $\sum_{i=1}^M h_n(\tilde{X}_i, Y_n, Z_n)/M$. The following proposition shows that a slight modification of this procedure is guaranteed to yield an \E-statistic.
	
	\begin{proposition} \label{prop:integral}
    Let $\tilde{X_1}, \dots, \tilde{X}_M$ be independent with distribution $Q_{Z_n}$. Then
    \[
        \check{E}_{h_n}^{\CI}(D_n):=\frac{h_n(X_n, Y_n , Z_n)}{(h_n(X_n,Y_n,Z_n) + \sum_{i=1}^M h_n(\tilde{X}_i, Y_n, Z_n))/(M+1)}
    \]
    satisfies $\mathbb{E}_P[\check{E}_{h_n}^{\CI}(D_n)\mid D^{n-1}] = 1$, $P \in \mathcal{H}_0$, where $h_n$ is as in Theorem~\ref{thm:e_var} and the expectation is taken both over the data and $\tilde{X_1}, \dots, \tilde{X}_M$.
	\end{proposition}
    
	In fact, the proof of Proposition \ref{prop:integral} only requires that $X_n, \tilde{X_1}, \dots, \tilde{X}_M$ are exchangeable, so it will also be applicable in many situations where data is randomly permuted.
	Furthermore, the naive approach without including $h_n(X_n,Y_n,Z_n)$ in the denominator is anti-conservative and does not define a sequence of conditional \E-statistics. Indeed, taking expectation over $X_n, \tilde{X}_1, \dots, \tilde{X}_M$,
    \begin{equation} \label{eq:naive_e}
        \mathbb{E}_P\left[\frac{h_n(X_n, Y_n , Z_n)}{\sum_{i=1}^M h_n(\tilde{X}_i, Y_n, Z_n)/M} \middle| Y_n, Z_n \right] \geq \frac{\int h_n(x,Y_n,Z_n) \, dQ_{Z_n}(x)}{\sum_{i=1}^M \int h_n(x,Y_n,Z_n) \, dQ_{Z_n}(x)/M} = 1,
    \end{equation}
	Here we use that $X_n, \tilde{X}_1, \dots, \tilde{X}_M$ are independent with distribution $Q_{Z_n}$, and invoke Jensen's inequality with the strictly convex function $s \mapsto 1/s$. Equality in \eqref{eq:naive_e} holds if any only if $h_n$ is constant in $x$, and in that trivial case the \E-statistic is equal to the constant $1$. 
	In all our applications, we use the variant proposed in Proposition \ref{prop:integral} to compute the \E-statistics.
	
	\subsection{Optimality} \label{sec:optimality}
	In order to accumulate evidence against the null hypothesis, the functions $h_i$ in~\eqref{eq:test_martingale} should measure the conditional (in)dependence between $X$ and $Y$. 
	This will ensure that the test martingale $(S_{h^n}^{\CI}(D^n))_{n\in\mathbb{N}}$ will grow if the null hypothesis is violated. 
	Ideally, we would be able to choose a measure of conditional independence that requires little to no assumptions on the distribution of $(X,Y,Z)$.
	Many such measures have been proposed in the literature, see e.g.~\citet{fukumizu2007kernelmeasures,Shah_2020,azadkia2021conddep}. 
    However, as far as we can tell, none of these measures allow for a sequential decomposition.
    That is, they are defined for fixed sample size $n$ as a function $T_n:\mathcal{D}^n\to [0,\infty)$, but cannot be decomposed in a nontrivial way as a product $T_n(D^n)=\prod_{i=1}^n T_i(D_i)$. 
    Our only option to use them in our test martingale is therefore to set $h_i(D_i)=T_i(D^{i-1},D_i)$ in \eqref{eq:test_martingale}. 
    This would have two major drawbacks: first, it would be computationally involved, because $T_i$ needs to be recalculated entirely \emph{within} the integral in the denominators in \eqref{eq:test_martingale}.
    Secondly, it would generally be ineffective, because $T_i(D^{i-1}, D_i)$ will depend very little on $D_i$ for large $i$, so $h_i(D_i)$ will generally not be sensitive to changing $X_i$ to $x'\sim Q_{Z_i}$. 
    As a result, the fraction in \eqref{eq:test_martingale} will be close to 1, preventing us from accumulating much evidence against the null.
    
    However, if we are willing to assume that under the alternative, the density of $(X,Y,Z)$ is given by $f$, then the conditional density $f_{Y\mid X,Z}$ itself is a measure of conditional independence. 
    Moreover, evaluating the density in $n$ data points is equivalent to taking the product of the density evaluated at all single data points $i=1,\dots,n$, because the data stream is i.i.d. 
    This gives a sequential decomposition as desired above.
    Furthermore, \citet{Katsevich2020OnModel-X} have shown that an optimal conditionally valid p-value based test can be achieved by running the CRT with the conditional density $f_{Y\mid X,Z}$ as test function.
	It turns out that this choice also yields the GRO \E-statistic among all \E-statistics defined on $n$ observations, and the expectation of the logarithm of this e-statistic is the conditional mutual information \citep{Cover1991}, an established conditional dependence measure which has also been applied for conditional independence testing \citep{Runge2018}.
	Note that the expectation is taken over the entirety of the data, i.e. $D^n=(X^n,Y^n,Z^n)$, as opposed to the result by~\citet{Katsevich2020OnModel-X} which only holds conditionally on $(Y^n,Z^n)$; see their article for a more thorough discussion.
	
	\begin{theorem}\label{thm:GRO}
		The GRO \E-statistic for testing $\mathcal{H}_0$ as in~\eqref{eq:null_hypothesis} against the alternative distribution with density $f$ is given by
		\begin{equation}\label{eq:e_information}
			S^{\CI}_{f_{Y\mid X,Z}}(D^n)=\prod_{i=1}^n E^{\CI}_{f_{Y\mid X,Z}}(D_i)=\prod_{i=1}^n \frac{f_{Y\mid X, Z}(Y_i \mid X_i, Z_i)}{f_{Y \mid Z}(Y_i\mid Z_i)}
		\end{equation} 
		and achieves growth rate
		\[
		\mathbb{E}_f[\log S^{\CI}_{f_{Y\mid X,Z}}(D^n)]=n I_f(X;Y\mid Z),
		\]
		where $I_f(X;Y\mid Z)$ denotes the conditional mutual information if $(X,Y,Z)$ follows the distribution with density $f$.
	\end{theorem}
	
	\begin{remark}
	A simple application of Bayes theorem allows one to rewrite~\eqref{eq:e_information} to $S_{f_{Y\mid X,Z}}^{\CI}(D^n) = \prod_{i=1}^n f_{X\mid Y, Z}(X_i \mid Y_i, Z_i)/f_{X \mid Z}(X_i\mid Z_i)$, which shows that the resulting test martingale is in fact a likelihood ratio process.
	That is, it is the ratio between the true density of $X$ given $(Y,Z)$ under the alternative and that under the null (whereas the density in the denominator of \eqref{eq:e_information} does not have to correspond to the data generating distribution). 
    The latter follows, because under the null $f_{X\mid Y,Z}$ is equal to $f_{X\mid Z}$, which we assume to be well-specified and equal the density of $Q_{Z_i}$.
	Hence the resulting test for a simple alternative hypothesis $f$ is in fact a generalization of the SPRT, where the distribution under the null hypothesis changes with the variable $Z_i$. 
	It depends on the application at hand which formulation, i.e.~\eqref{eq:e_information} or conditional densities of $X$, is more suitable for constructing a test. 
	For example, we show in Section \ref{sec:logistic} that for binary $Y \in \{0,1\}$ and $(X, Z) \in \mathbb{R}^p \times \mathbb{R}^q$, one can construct a test based on logistic regression, which is often simpler than directly trying to find a suitable conditional density of $X$ given $Y$ and $Z$, especially when $p > 1$.
	\end{remark}
	
	The information inequality \citep{Cover1991} implies that $I_f(X;Y\mid Z) \geq 0,$
	with equality if and only if $Y \ind X \mid Z$, which shows that $ S^{\CI}_{f_{Y\mid X,Z}}$ has nontrivial power to detect deviations from conditional independence if $f$ is the true density of $(X,Y,Z)$.
	Assuming a simple (point) alternative $f$ is, of course, an unrealistically strong assumption. We now proceed to develop a method that also gets uniform growth rates for potentially large classes of alternative densities ${\cal F}$ by building on the simple alternative case. 
	The following result states that if we do not know the density $f_{Y\mid X,Z}$, but instead use a different density $g_{Y\mid X,Z}$, the loss in expected growth rate is directly related to a measure of distance between $f_{Y\mid X,Z}$ and $g_{Y\mid X,Z}$.
	
	\begin{proposition}\label{prop:misspec}
		For any conditional density $g_{Y\mid X,Z}$, the following holds:
		\begin{equation}\label{eq:misspec}
			\mathbb{E}_f\left[\log E^{\CI}_{g_{Y\mid X,Z}}(D)\right]\geq I_f(X;Y\mid Z)-\mathbb{E}_f[\mathrm{KL}(f_{Y\mid X,Z}\| g_{Y\mid X,Z})].
		\end{equation} 
	\end{proposition}
	
	Since we are in a sequential setting, this proposition implies that if we do not know the density $f$, we could try to estimate it, using estimates that improve as sample size increases.
	That is, let $\hat f_n$ be an estimator of $f_{Y\mid X,Z}$ based on data $D^{n}$, and $\hat f_0$ an initial guess.
	The test martingale we use is then given by $\prod_{i=1}^n E^{\CI}_{\hat{f}_{n-1}}(D^i)$.
	It follows from the combination of Theorem~\ref{thm:GRO} and Proposition~\ref{prop:misspec} that if the estimator is consistent in a KL sense, then  the expected growth per outcome converges to that of the GRO \E-statistic. 
	
	\begin{corollary}\label{cor:asympgro}\leavevmode
	\begin{enumerate}[label=(\roman*)]
	    \item Assume that $\frac 1n \sum_{i=1}^n \mathbb{E}_f[\mathrm{KL}(f_{Y\mid X,Z}\| \hat f_{i-1}) \mid D^{i-1}]\xrightarrow[n\to \infty]{\text{a.s.}}0$,
		then 
		\[\frac 1n \sum_{i=1}^n \mathbb{E}_{f}\left[\log E^{\CI}_{\hat f_{i-1}}(D_i) \big\vert D^{i-1}\right]\xrightarrow[n\to \infty]{\text{a.s.}} I_f(Y;X\mid Z).\]
		\item Assume that for some function $b(n): {\mathbb N} \rightarrow {\mathbb R}^+_0$ with $b(n) = o(n)$, we have
		\begin{equation}\label{eq:cii}
		\mathbb{E}_{f}\left[\sum_{i=1}^n \mathbb{E}_f [\mathrm{KL}(f_{Y\mid X,Z} \| \hat f_{i-1}) \mid D^{i-1}]\right]\leq b(n).
		\end{equation} 
		Then
		\begin{equation}	\label{eq:uniform}
  \frac1n \mathbb{E}_{f} \left[\sum_{i=1}^n \log E^\CI_{\hat f_{i-1}}(D_i) \right]\geq I_f(X;Y\mid   Z)-\frac{b(n)}{n}.\end{equation}
	\end{enumerate}
	\end{corollary}
	Consequently, to achieve an asymptotic optimal growth rate, we need to use a conditional density estimator $\hat{f}_i$ that converges in KL divergence to $f$, where we may assume $ f\in {\cal F}$ for some given set of densities ${\cal F}$, i.e. our statistical model. 
		\cite{Barron98} showed that, for a wide variety of parametric and nonparametric models ${\cal F}$, we  have {\em convergence in information\/} (his terminology for (\ref{eq:cii}) holding for all $n$)  if we set  $\hat{f}_i$ to be the Bayes predictive density, under no further conditions, uniformly for all  $f \in {\cal F}$, as long as a suitable prior is used. 
 This means that, for some fixed function $b$ (\ref{eq:cii}) holds uniformly for all $f \in {\cal F}$, so that also (\ref{eq:uniform}) holds uniformly for all $f \in {\cal F}$: we could put a $\inf_{f \in {\cal F}}$ to the left of (\ref{eq:uniform}) and the result would still hold. Thus, we get the optimal growth rate (which itself can only be achieved by an oracle that knows the `true' $f \in {\cal F}$) up to an additive term of $b(n)/n$ uniformly for all $f \in {\cal F}$. 
 The rate $b(n)/n$ is then usually, up to log factors, equal to the minimax rate in squared Hellinger distance. 
	The same rates (potentially up to further log factors) are available for the Bayesian posterior mean under a weak additional condition on the model introduced by \cite{Grunwald2020} under the name {\em witness-of-badness\/}; it generalizes a well-known earlier condition of \cite{WongShen95}; see also \citep{bilodeau2021minimax} for related results. 
 In Section~\ref{sec:logistic}, we demonstrate our approach with ${\cal F}$ set to the logistic regression model with $X \in {\mathbb R}^p$ and $Z \in {\mathbb R}^q$, for which a result by \cite{foster2018logistic} in combination with \citep{Barron98} implies that, if we use the Bayes predictive distribution as above, then $b(n)$ can be chosen as $O((p+q) \log n)$ as long as the first four moments of all components of $X$ and $Z$ exist, implying a parametric rate. However, in our experiments in Section~\ref{sec:sim}, rather than the Bayes predictive distribution, we use a (slightly regularized) MLE, since it can be computed much more efficiently. Although we suspect that the MLE converges at the same rates as Bayesian methods under the same weak conditions, the methods of the aforementioned papers cannot be used to prove this, and instead in Proposition~\ref{prop:logistic} we show almost sure convergence of the MLE, without rates, under a stronger subgaussianity assumption on $(X,Z)$.

	\subsection{Worst-Case Bounds on Rejection Rate}\label{sec:rej_bound}
    Up to now, we have discussed the construction and properties of \E-statistics when the conditional distributions $Q_z$ are known exactly.  
	In this section, we prove a result analogous to Theorem 4 of \citet{Berrett2020} 
	(see \eqref{eq:crt_robustness})
	on the error rate of our sequential test under the null hypothesis when the distributions $Q_z$ are only approximations.
	The approximation of $Q_z$ will be denoted by $\hat{Q}_z$, and the (approximate) \E-statistic at time $n$ is given by
	\[
	\tilde{E}^{\CI}_{h_n}(D^n) = \frac{h_n(X_n, Y_n, Z_n \mid D^{n-1})}{\int_{\mathcal{X}}h_n(x, Y_n, Z_n \mid D^{n-1}) \, \mathrm{d}\hat{Q}_{Z_n}(x)}.
	\]
	Here the nonnegative function $h_n$ depends on $D^{n-1}$ since it can be constructed sequentially, for example by estimating the conditional density of $Y_n$ given $X_n$ and $Z_n$ with all past data $(X_i, Y_i, Z_i)$, $i = 1, \dots, n - 1$. Recall that $Q^n_{Z^n}$ denotes the product distribution of $Q_{Z_i}$, $i = 1, \dots, n$, that is, for measurable $A \subseteq \mathcal{X}^n$
	\[
	Q^n_{Z^n}(A) = \int_{\mathcal{X}} \dots \int_{\mathcal{X}} 1\{x^n \in A\} \, \mathrm{d}Q_{Z_1}(x_1) \cdots \mathrm{d}Q_{Z_n}(x_n),
	\]
	In particular, $Q^n_{Z^n}(A) = P(X^n \in A \mid Z^n) =P(X^n \in A \mid Y^n, Z^n)$ for $P \in \mathcal{H}_0$, due to conditional independence of $Y_i$ and $X_i$ given $Z_i$, $i = 1, \dots, n$. The distribution $\hat{Q}^n_{Z^n}$ is defined in the same way as $Q^n_{Z^n}$ but with $\hat{Q}_{Z_i}$ instead of $Q_{Z_i}$.
	\begin{theorem} \label{thm:bound}
		Assume that $h_1, \dots, h_N > 0$ are measurable. For any $N \in \mathbb{N}$, $\alpha \in (0,1)$ and $P \in \mathcal{H}_0$,
		\begin{equation}
		    \label{eq:VilleIsOnTV}
		P\left(\exists \, n \leq N \colon \prod_{i = 1}^n \tilde{E}^{\CI}_{h_i}(D^i) \geq \frac{1}{\alpha}  \middle| Y^N, Z^N \right) \leq \alpha + d_{\mathrm{TV}}(Q^N_{Z^N}, \hat{Q}^N_{Z^N}).
			\end{equation}
	\end{theorem}
	
	Theorem \ref{thm:bound} gives the same worst case bound on the rejection rate as Theorem 4 in \citet{Berrett2020} for a sample size $N$, but optional stopping at any sample size $n \leq N$ is allowed. \citet[Section 5.1]{Berrett2020} discuss conditions under which the total variation distance between $Q^N_{Z^N}$ and $\hat{Q}^N_{Z^N}$, which bounds the excess rejection rate, is small. For example, they obtain an upper bound of the form $\mathcal{O}_p(\sqrt{Nk/m})$ when $(X, Z) \in \mathbb{R}^k$ follow a multivariate Gaussian distribution and the conditional law of $X \mid Z$ is estimated with an unlabeled sample of size $m$. Hence when $m$ remains constant but $N$ diverges to infinity, the bound on the distance between $Q^N_{Z^N}$ and $\hat{Q}^N_{Z^N}$ becomes trivial.
	Furthermore, the discussion at the end of Section~\ref{sec:optimality} on estimation in terms of KL divergence (which bounds the total variation distance) can also be applied to the estimation of $Q_z$.
	
	A natural question is whether in the sequential setting, one could continuously update the approximation of the conditional distributions of $X\mid Z$ at stage $n$ with the new data $X_i, Z_i$, $i = 1, \dots, n - 1$, obtained so far, similarly in spirit to the estimator $\hat{f}_n$ we mentioned above Corollary~\ref{cor:asympgro}. This would mean that at time $n$, the conditional distribution $\hat{Q}_{Z_n}$ also depends on $X_i, Z_i$, $i = 1, \dots, n - 1$, and we denote it as $\bar{Q}_{Z^n,X^{n-1}}$. In this case, a similar bound as in Theorem \ref{thm:bound} indeed applies. Let $\bar{Q}^N_{Z^N}$ be the distribution constructed as follows. Generate $X_1\mid Z_1 \sim \hat{Q}_{Z_1}$, and then, for $i = 2, \dots, N$,
	\[
	X_i \mid Z^{i}, X^{i-1} \ \sim \ \bar{Q}_{Z^i, X^{i-1}}.
	\]
	Hence, different from $\hat{Q}^N_{Z^N}$, the random variables $X_1, \dots, X_N$ are not independent under $\bar{Q}_N$. But with exactly the same proof as for Theorem \ref{thm:bound}, one may obtain that the excess rejection rate is again bounded by $d_{\mathrm{TV}}(Q^N_{Z^N}, \bar{Q}^N_{Z^N})$. While this is qualitatively the same bound, it is an open question whether one can find settings in which this upper bound is smaller (so that perhaps the bound (\ref{eq:VilleIsOnTV}) would not become trivial any more for large $N$).
	
	\subsection{Application to Logistic Regression} \label{sec:logistic}
	The general construction strategy for \E-statistics and all results so far assume no specific model for the outcome $Y$ or covariates $(X,Z)$. In this section, we consider the special but important case of a binary outcome $Y \in \{0,1\}$ which  follows a logistic regression model under the alternative, i.e.~$(X,Z) \in \mathbb{R}^{p+q}$, and $Y$ equals $y = 0, 1$ with probability
	\begin{equation} \label{eq:logistic}
		p_{\theta}(y \mid X, Z) = \frac{\exp(y(\beta^\top X + \gamma^\top Z))}{1+\exp(\beta^\top X + \gamma^\top Z)},
	\end{equation}
	with an unknown $(p+q)$-dimensional coefficient vector $\theta = (\beta_1, \dots, \beta_p, \gamma_1, \dots, \gamma_q)$. Conditional independence of $Y$ and $X$ given $Z$ holds if and only if $\beta_1 = \dots = \beta_p = 0$. It turns out that in this setting, one can construct \E-statistics which not only have power on average, as in Corollary \ref{cor:asympgro}, but which even reject the null hypothesis with probability one if it is violated and the sample size grows to infinity. At this point, it is important to recall that the validity of the \E-statistics does not require the logistic model to be correctly specified: the validity only depends on the null hypothesis, which is still the set of {\em all\/} distributions under which conditional independence holds in the sense of (\ref{eq:null_hypothesis}), including many distributions that violate the logistic model assumption. The result rather shows that {\em if\/} the logistic model is suitable (in the sense that, if the alternative is true, then data are sampled from a distribution in this model), then the \E-statistic has guaranteed power to detect violations of conditional independence. 
	
	Following our general strategy, an \E-statistic for testing conditional independence is given by
	\begin{equation} \label{eq:logistic_evalue}
		S^{CI}_n(D^n) = \prod_{i=1}^n \frac{p_{\hat{\theta}_{i-1}}(Y_i \mid X_i, Z_i)}{\int p_{\hat{\theta}_{i-1}}(Y_i \mid x, Z_i) \, \mathrm{d}Q_{Z_i}(x)},
	\end{equation}
	where $\hat{\theta}_{k}$ may be any estimator for $\theta$ based on the first $k$ samples, $(X_i, Y_i, Z_i)$, $i = 1, \dots, k$. When the observations $(X_i, Y_i, Z_i)$, $i \in \mathbb{N}$, are independent and identically distributed, the growth rate optimal \E-statistic is obtained if $\hat{\theta}_{k} = \theta$ for all $k \in \mathbb{N}$. 
	Nevertheless, the following proposition shows that tests based on $S^{CI}_n$ have asymptotic power one when $\hat{\theta}_k$ is the maximum likelihood estimator. From now on, $\|v\| = (v^{\top}v)^{1/2}$ denotes the Euclidean norm of a vector $v$, and we denote by $(X,Z)$ the stacked vector $(X_1, \dots, X_p, Z_1, \dots, Z_q)$. 
	We will take $(X, Y, Z)$ as a generic observation which has the same distribution as $(X_i, Y_i, Z_i)$, $i \in \mathbb{N}$, for writing probability statements about elements of this sequence.
	
	\begin{proposition} \label{prop:logistic}
		Let $(X_i, Y_i, Z_i)$, $i \in \mathbb{N}$, be independent and identically distributed such that \eqref{eq:logistic} holds with $(\beta_1, \dots, \beta_p) \neq 0$. Assume furthermore that
		\begin{enumerate}[label=(\roman*)]
			\item (a) $(X,Z)$ satisfies $P(u^{\top}(X,Z) \neq 0) > 0$ for all $u \in \mathbb{R}^{p+q} \setminus \{0\}$, and (b) it is subgaussian with variance parameter $\sigma^2$, that is
			\[
			\mathbb{E}[\exp(u^{\top}((X,Z)-\mathbb{E}[(X,Z)]))] \leq \exp(\|u\|^2\sigma^2/2), \ \forall \, u \in \mathbb{R}^{p+q},
			\] 
			\item $\hat{\theta}_{n}$ in \eqref{eq:logistic_evalue} is the logistic MLE based on data $(X_i, Y_i, Z_i)$, $i = 1, \dots, n$, for all $n \in \mathbb{N}$, with $\hat{\theta}_{n}$ arbitrarily defined but finite if the MLE does not exist. 
		\end{enumerate}
		Then $S_n^{CI}$ satisfies $\liminf_{n \rightarrow \infty} \log(S_n^{CI})/n \geq I(X;Y|Z) > 0$ almost surely.
	\end{proposition}
	
	Assumption (i)(a) ensures that the MLE converges almost surely at a fast rate, as shown by \citet{Qian2002}. Instead of subgaussianity ((i)(b)) their result requires only moment assumptions (which are implied by subgaussianity), but subgaussianity is indeed required in our setting: see the proof of Proposition \ref{prop:logistic}, which is given in Appendix \ref{app:proof_logistic}.
	
	\section{Simulations} \label{sec:sim}
	To investigate the robustness and power of tests of conditional independence based on \E-statistics, we compare the \E-statistic \eqref{eq:logistic_evalue} for binary $Y$ to other methods applicable in this setting.
	The covariate $X$ is univariate, $X \in \mathbb{R}$, while $Z = (1, Z_1, \dots, Z_{q-1})$ is a $q$-dimensional vector containing an intercept term.
   The distribution of $(X, Z_1, \dots, Z_{q-1})$ is the $q$-dimensional normal distribution with zero mean and a Toeplitz covariance matrix, $\Sigma_{i,j} = 1/(1+|i-j|)$ for $i, j = 1, \dots, q$. Then $Q_Z$ is the Gaussian distribution with mean
	\begin{equation} \label{eq:true_conditional_mean}
		\mu_Z =  \Sigma_{1, -1} \Sigma_{-1, -1}^{-1} (Z_1, \dots, Z_{q-1})^{\top}
	\end{equation}
	where $\Sigma_{-1, -1}$ is the submatrix $(\Sigma_{i,j})_{i,j=2, \dots, q}$ and $\Sigma_{1,-1}$ is the row vector $(\Sigma_{1, 2}, \dots, \Sigma_{1,q})$. The conditional variance equals $\sigma_Z^2=1 -\Sigma_{1, -1} \Sigma_{-1, -1}^{-1} \Sigma_{1,-1}^{\top}$. The binary response $Y$ has probabilities given by $p_{\theta}(y \mid X, Z)$, $y \in \{0,1\}$, as in~\eqref{eq:logistic}, where the intercept and the coefficients of $Z$, and $\gamma_1, \dots, \gamma_q$, are drawn independently and uniformly distributed on the interval $[-1, 1]$. The coefficient of $X$, i.e.~$\beta = \beta_1$, is chosen in $[0,1]$, with $0$ corresponding to conditional independence of $Y$ and $X$ given $Z$. 
	Below are implementation details for all methods considered in the simulations.
	In the following subsections, these methods are compared in terms of type-I error and power. 
	A further study on the robustness of the MX-based methods with respect to violations of the MX assumption is given in Appendix \ref{sec:robustness}.
	
	\paragraph{Conditional randomization \E-statistic (E-CRT).} The parameter vector $\theta$ in \eqref{eq:logistic_evalue} is re-estimated after each new observation with the maximum likelihood method, starting from a minimal sample size of $5q + 1$, so that $5q$ observations are available for the first parameter estimate. In addition, the probabilities $p_{\hat{\theta}_k}(y \mid X, Z)$ are truncated to $[\varepsilon, 1 -\varepsilon]$ for some small $\varepsilon > 0$. This is to account for the fact that at small sample sizes the MLE sometimes yields predicted probabilities in $\{0,1\}$. We also include an oracle version of this \E-statistic (E-CRT-O), which uses the true $\theta$ starting from the first observation, instead of the maximum likelihood estimator. For both variants, the integral in the \E-statistic is approximated by an average over $500$ Monte Carlo samples.
	
	\paragraph{Conditional randomization test (CRT).} The CRT is applied nonsequentially with the likelihood of the logistic regression model as test statistic and $500$ samples for randomization. That is, $X$ is sampled $500$ times from the conditional distribution given $Z$, the logistic regression model is re-estimated with this simulated covariate, and the likelihood achieved with these models with simulated $X$ is compared to the likelihood achieved with the actual values of $X$.
	
	\medskip
	The following methods are for testing whether the coefficient $\beta$ in the logistic regression model equals zero. Unlike the E-CRT and CRT, they are not based on the MX assumption, but their type-I error guarantee does require that the true probabilities of $Y$ are given by the logistic model \eqref{eq:logistic}.
	A comparison is of interest since these methods are, to the best of our knowledge, currently the only ones that allow sequential testing in a logistic model.
	
	\paragraph{Running maximum likelihood (R-MLE).} We apply the running MLE method by \citet[Section 7]{Wasserman2020},
	an instance of the generic method introduced in that paper which they call {\em universal inference}.
	Parameter estimation is also started with a minimum $5q$ observations, and we additionally investigate an $L_1$ penalized version for estimation under the alternative hypothesis, abbreviated as R-MLE-P, which, like for the E-CRT, is to prevent predicted probabilities close to $0$ or $1$ due to divergence of the MLE. In this second variant, the penalization parameter is chosen by $10$-fold cross validation on the available data at the given time, with likelihood as optimization criterion. The penalization parameter is only updated every $10$ observations, since the cross validation is computationally expensive.
	
	\paragraph{Likelihood ratio test (LRT).} The classical asymptotic likelihood ratio test for the null hypothesis that $\beta = 0$ is applied with fixed sample size, and group sequential versions of it with $K$ equally sized groups and the methods by \citet{Pocock1977} (LRT-PK) and \citet{Obrien1979}(LRT-OF).
	
	\medskip
	The results shown in this section are for dimension $q = 4$ of the covariate vector $(X,Z)$. 
	A maximum sample size of $n = 2000$ is considered, after which the evaluation is terminated independently of whether the null hypothesis is rejected. The sequential methods apply the most aggressive stopping rule, namely, reject the null hypothesis as soon as the test statistic exceeds $1/\alpha$ once; more discussion on this is provided in Section \ref{sec:sim_alt}. All results, i.e.~rejection rates and average sample sizes, are computed over $800$ simulations of the data generating process. The same simulation but with higher dimension ($q = 8$) or with negative correlations between the covariates ($\Sigma_{i,j} = (-1)^{i-j}/(1+|i-j|)$) yields similar results. For the running MLE method, we additionally tested whether not penalizing the coefficient of interest, $\beta$, may achieve higher power, which was not the case.
	
	The simulation is performed in \textsf{R} 4.2 \citep{Rcore2022}, with the {\tt glm} function for maximum likelihood estimation of the logistic regression parameter, the package {\tt glmnet} \citep{Simon2011} for $L_1$ penalized estimation, and the package {\tt ldbounds} \citep{Casper2022} for computing critical values for the Pocock and O'Brien-Fleming group sequential tests.
	\if1\blind { Replication material for the simulations and the case study, as well as additional figures, are available on \url{https://github.com/AlexanderHenzi/eindependence}. } \fi
	
	\subsection{Sequential Tests Under the Null} \label{sec:sim_null}
	Table \ref{tab:sim_null} shows the rejection rates of the different sequential methods with a maximum sample size of $n = 2000$. 
	The methods based on \E-statistics and the running MLE yield rejection rates below the nominal level. For the \E-statistics, the chosen truncation level is $\varepsilon = 0.05$, but the rejection rate is also below $\alpha$ for $\varepsilon = 0,\, 0.01,\, 0.1$. The group sequential methods with $K = 20$ equally sized groups, each of size $100$, are slightly anti-conservative, and similar rejection rates are obtained for $K = 5, \, 10, \, 40$.
	
	\begin{table}[ht]
		\centering
		\resizebox{\textwidth}{!}{
		\begin{tabular}{*{7}{c}}
			\toprule
			&   E-CRT    &   R-MLE & R-MLE-P    &    LRT-PK   & LRT-OF \\
			\midrule
			$\alpha = 0.01$ &   0.0075 (0.0031)   &   0.0038 (0.0022)   &   0.0038 (0.0022)   &   0.0138 (0.0013)   &   0.0127 (0.0040)   &  \\
			\midrule
			$\alpha = 0.05$ &   0.0438 (0.0072)   &   0.0038 (0.0022)   &   0.0038 (0.0022)  &   0.0700 (0.0090)   &   0.0599 (0.0084)   &  \\
			\bottomrule
		\end{tabular}
		}
		\caption{Rejection frequencies (and standard errors) of the different methods, with implementation details as described in Section \ref{sec:sim_null}. Frequencies are given in $[0,1]$, not in percentages.}
		\label{tab:sim_null}
	\end{table}
	
	\subsection{Simulations Under the Alternative}\label{sec:sim_alt}
	We proceed to compare the different methods under violations of the null. 
	This is commonly done by comparing the achieved power at given sample sizes $n$ for different effect sizes $\beta$, or the inverse of that function, i.e.~the minimum sample size required to achieve a certain power $1-\eta$,
	\[
	    N(\beta, \eta) = \min\{n \in \mathbb{N}\colon P_{\beta}(\phi_n = 1) \geq 1-\eta\}.
	\]
	For a fixed Type-I error probability $\alpha$, the test decision $\phi_n$ is given by $\phi_n = 1\{\max_{m \leq n} S_m \geq 1/\alpha\}$ for anytime-valid tests based on $(S_n)_{n\in\mathbb{N}}$, or $\phi_n = 1\{p_n \leq \alpha\}$ for a method based on a fixed sample size p-value $p_n$. For an anytime-valid test, $N(\beta,\eta)$ can be regarded as the worst case sample size a researcher has to plan for in order to achieve power $1-\eta$; the actual sample size at rejection may be smaller thanks to optional stopping.
	Therefore, we additionally consider the average sample size of the anytime-valid tests when evaluation is terminated at the latest at $N(\beta, \eta)$,
	\[
	    N_{\mathrm{av}}(\beta,\eta) = \mathbb{E}_{P_{\beta}}[\min(N(\beta,\eta), \, \inf\{n \in \mathbb{N}\colon S_n \geq 1/\alpha\})].
	\]
	The rationale is that --- even though we have seen that anytime-valid methods retain Type-I error validity under arbitrary stopping times --- in practice, a natural way to proceed with an anytime-valid test is to run the experiment until either a rejection or a given upper bound on the number of samples is reached.
	The obvious choice for this upper bound is $N(\beta,\eta)$, as it ensures that the test will have a power of $1-\eta$.
	Then $N_{\mathrm{av}}(\beta,\eta)$ gives the average sample size of tests conducted with an anytime-valid method designed to have power $1-\eta$. 
	
\begin{figure}[ht]
	    \begin{center}
	    \includegraphics[width=0.8\textwidth]{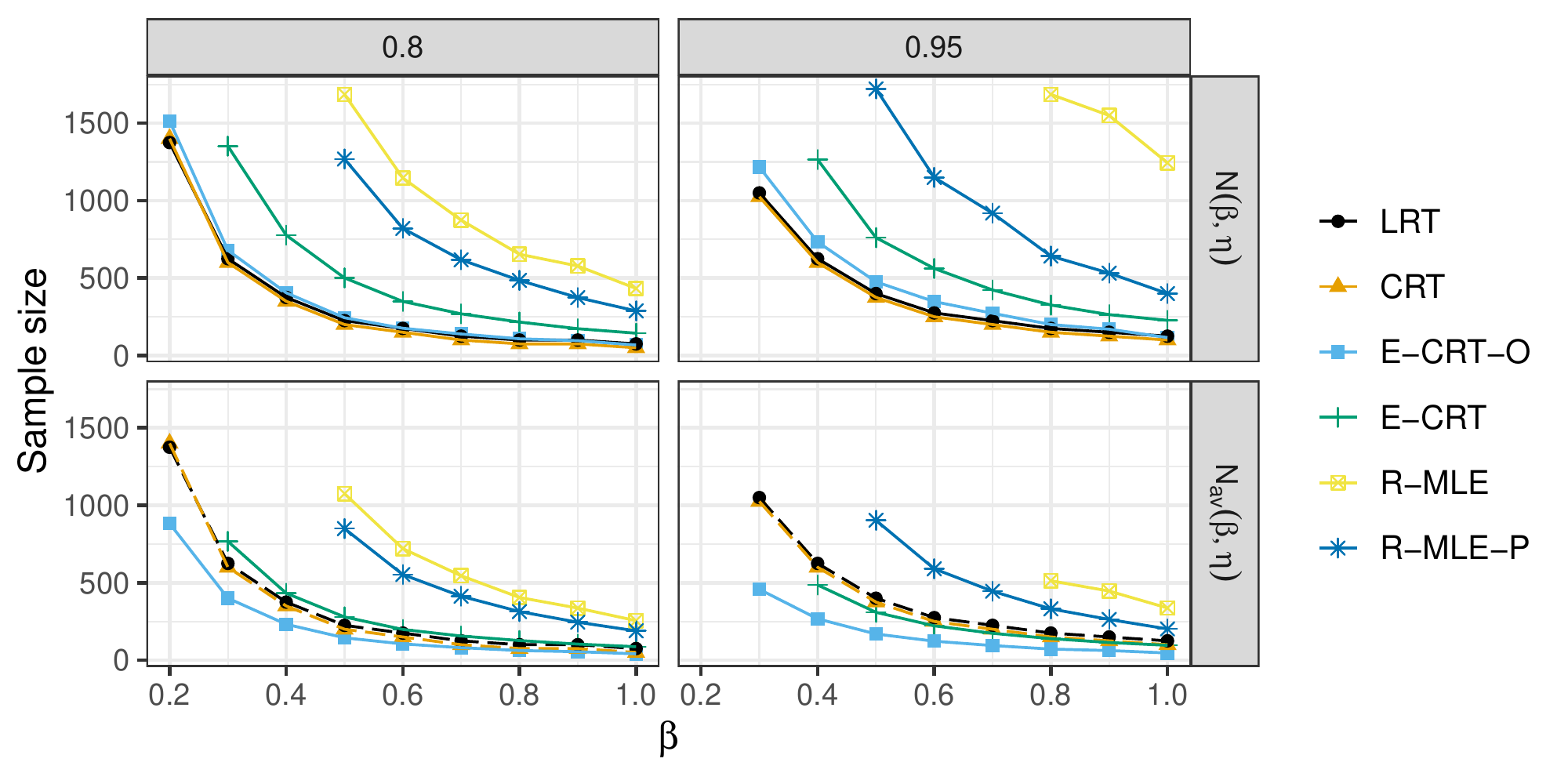}
		\caption{
		Sample sizes for the different testing methods to achieve power $1-\eta = 0.8, \, 0.95$ with Type-I error $0.05$. For the nonsequential methods (LRT, CRT), dashed lines show $N(\beta,\eta)$ also in the lower panels for better comparison. Simulations were conducted only up to $n = 2000$, and no results are shown if $N(\beta,\eta) > 2000$ for a given method, $\beta$ and $\eta$, i.e.~if more than $2000$ observations would be required to achieve a power of $1-\eta$.
		\label{fig:alternative}}
	    \end{center}
	\end{figure}

	A comparison of the different methods in terms of $N(\beta,\eta)$ and $N_{\mathrm{av}}(\beta,\eta)$ is given in Figure~\ref{fig:alternative}.
	The group sequential methods are excluded from this figure and are analyzed in more detail at the end of the section. 
	The upper two panels of Figure~\ref{fig:alternative} depict $N(\beta,\eta)$ for $1-\eta = 0.8, \, 0.95$ as a function of the parameter $\beta$ (for clarity note that, as before, $\beta$ stands for the parameter vector in \eqref{eq:logistic_evalue}; we never use $1-\beta$ for power). 
	It can be seen that $N(\beta, \eta)$ is higher for the anytime-valid methods than for fixed sample size tests. 
	This is to be expected: the sample size $N(\beta, \eta)$ ensures power $1-\eta$, but the actual sample size of an anytime-valid test is random and often smaller thanks to early stopping. 
	The lower two panels of Figure~\ref{fig:alternative} similarly show $N_{\mathrm{av}}(\beta,\eta)$ as a function of $\beta$. 
	It can be seen that the average sample size is not more and sometimes even less than the sample size of the nonsequential tests. 
	These results suggest that the average sample size to reject the null hypothesis with the E-CRT is not higher than the fixed sample size one would have to plan for with the CRT or LRT. 
	Similar observations have already been made for \E-statistics in other settings, such as comparisons with the t-test \citep{Gruenwald2019}, Fisher's Exact Test \citep{Turner2021}, or the logrank test \citep{terSchure2021}. 
	
    Furthermore, the running MLE method requires much more data to achieve a rejection than the other methods, even with the superior penalized estimation under the null hypothesis. 
    The large sample sizes required for the R-MLE to have a power of $0.95$ are mainly due to predicted probabilities close or equal to zero or one at early stages, a problem which is remedied by using penalization, as already proposed by \citet{Wasserman2020}; even then, more data are required though. 
    We again emphasize that the running MLE methods are based on different assumptions than the randomization based tests: they do not require the MX assumption, but are only valid for a correctly specified logistic model.
    However, even when compared to the classical likelihood ratio test, which requires almost the same sample sizes as the CRT, one would have to plan for substantially higher sample sizes with the running MLE methods.

	For the conditional randomization \E-statistics, we tested different levels of truncation $[\varepsilon, 1-\varepsilon]$ for the predicted probabilities. Truncating at a small level $\varepsilon = 0.01, \, 0.05$ is superior to no truncation, $\varepsilon = 0$, since it prevents \E-values close to or equal to zero, but if the truncation level becomes too high, $\varepsilon = 0.1$, it limits the power of the test for observations with predicted probability close to zero or one. See Figure \ref{fig:alternative_eps0} in Appendix \ref{app:extra_sim} for the case $\varepsilon = 0$. The results shown in this and the next section are for $\varepsilon = 0.05$. In principle, one could also apply a penalized estimator to remedy convergence problems of the MLE, but truncation is computationally less demanding as it does not require the selection of a penalization parameter.

\begin{figure}[ht]
	    \centering
		\includegraphics[width=\textwidth]{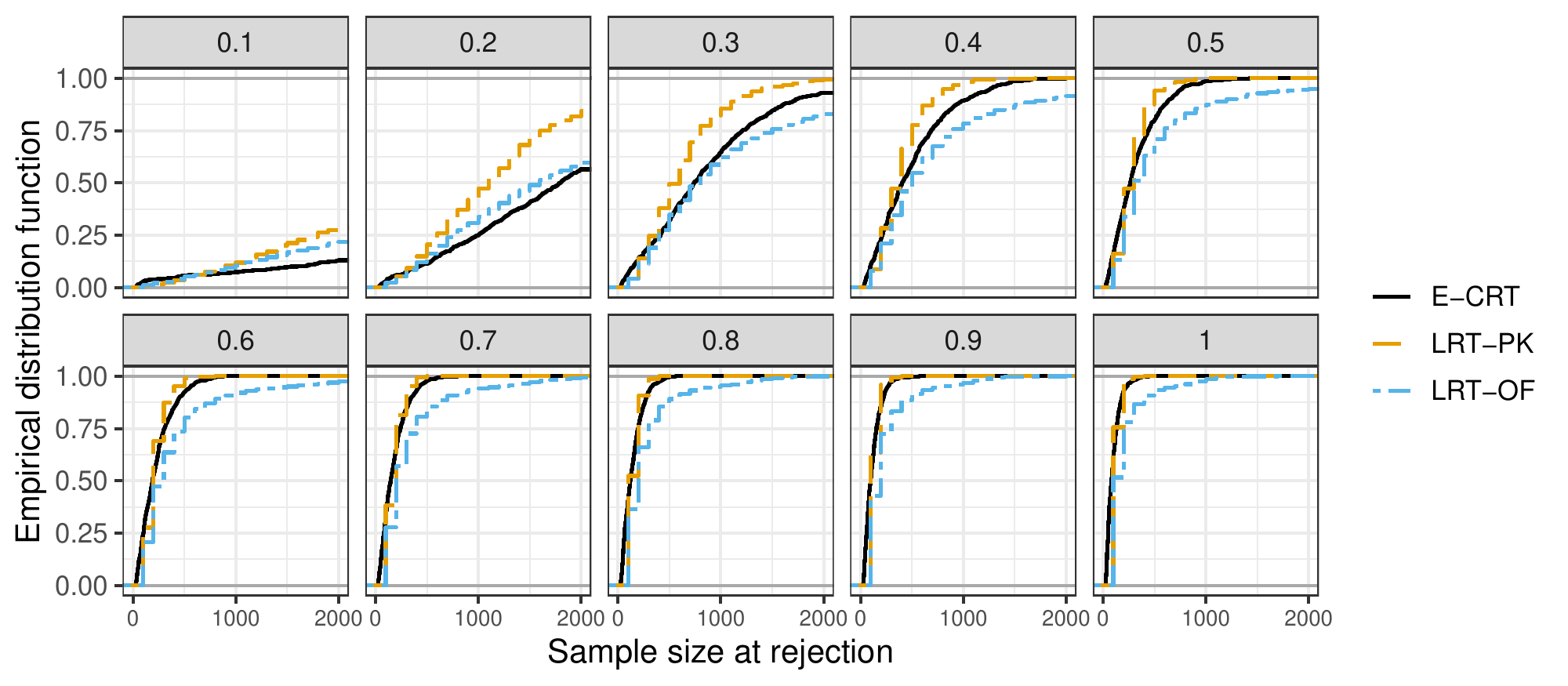}
		\caption{Empirical distribution of the sample size at rejection, with level $\alpha = 0.05$, for the randomization based \E-statistics and group sequential methods, and $\beta \in \{0.1, 0.2, \dots, 1\}$. \label{fig:group_sequential}}
	\end{figure}

	Finally, in Figure \ref{fig:group_sequential}, the conditional randomization \E-statistic is compared to group sequential methods with the Pocock and the O'Brien and Fleming method with $K = 20$ groups. We see that for small parameter $\beta \in \{0.1, 0.2\}$ these methods achieve a higher power than the \E-statistics, but as already shown in Table \ref{tab:sim_null}, the group sequential methods do not control the rejection rate below the nominal level when $\beta = 0$. Also, the E-CRT yields slightly more rejections at small sample sizes than the group sequential methods. As $\beta$ increases, the conditional randomization \E-statistics tend to outperform the O'Brien and Fleming method, and achieve a rejection rate very close to Pocock's method. Even for large $\beta$, the O'Brien and Fleming requires higher sample sizes due to the fact that the method is designed to yield fewer rejections with small samples. Different numbers of groups for the group sequential methods give similar results, except for the fact that rejecting at small sample sizes becomes impossible if the number of groups is small and the group size large. The performance of the \E-statistics compared to the group sequential methods is in line with the results of \citet{terSchure2021} for survival analysis. Also in their study, group sequential methods and alpha-spending approaches, which have to stop at a certain maximum sample size, tend to achieve a higher power than open-ended tests based on \E-statistics, which do not require a finite upper bound on the sample size.
	
	\section{Data Application}\label{sec:data}
	\citet{Berrett2020} analyze conditional independence relations in the Capital Bikeshare data set. Code and data for their study are available on \url{https://rinafb.github.io/research/, https://ride.capitalbikeshare.com/system-data}. The data set collects information on bike rides with the Capital bikeshare System in Washington DC. One question in the analysis by \cite{Berrett2020} is whether there is dependence of the duration of ride and the binary variable member type, distinguishing casual users from people who purchased a long-term membership, conditional on the start location, destination, and the time of the day during which the bike ride took place.
	
	Let $X$ denote the logarithm of the ride duration and denote by $Z$ the three dimensional vector of starting point, destination of ride, and of the starting time. 
	We model the distribution of $X \mid Z$ as Gaussian, $\mathcal{N}(\mu_{Z}, \sigma_{Z}^2)$ in exactly the same way as \citet{Berrett2020}. Separately for each combination of starting point and destination, the conditional mean $\mu_{Z}$ and variance $\sigma_{Z}^2$ are estimated by a kernel regression of $X$ with the time of day as covariate; details are given in Appendix B of \citet{Berrett2020}. We also apply the same preselection criteria, namely, exclude data from weekends and holidays, and values of $Z$ where the estimation might be imprecise due to scarce data. The member type is encoded as $Y \in \{0,1\}$, with $Y = 0$ referring to casual members.
	
	The data analysed is from the months September to November in 2011. Unlike \citet{Berrett2020}, who took the months September and November as training data for estimating the distribution of $X\mid Z$ and October as test data, we perform estimation on data of September and October and perform tests on the November data, which would be the natural order for real-time sequential analysis. The test data set, after the application of selection criteria, contains $7173$ observations. The training data consists of $158\,741$ observations. Due to the temporal structure of the data there might be some short lag autocorrelation between the observations, but we did not find any distorting influence of this on the results presented below.
	
\begin{figure}[ht]
	\begin{center}
		\includegraphics[width=.75\textwidth]{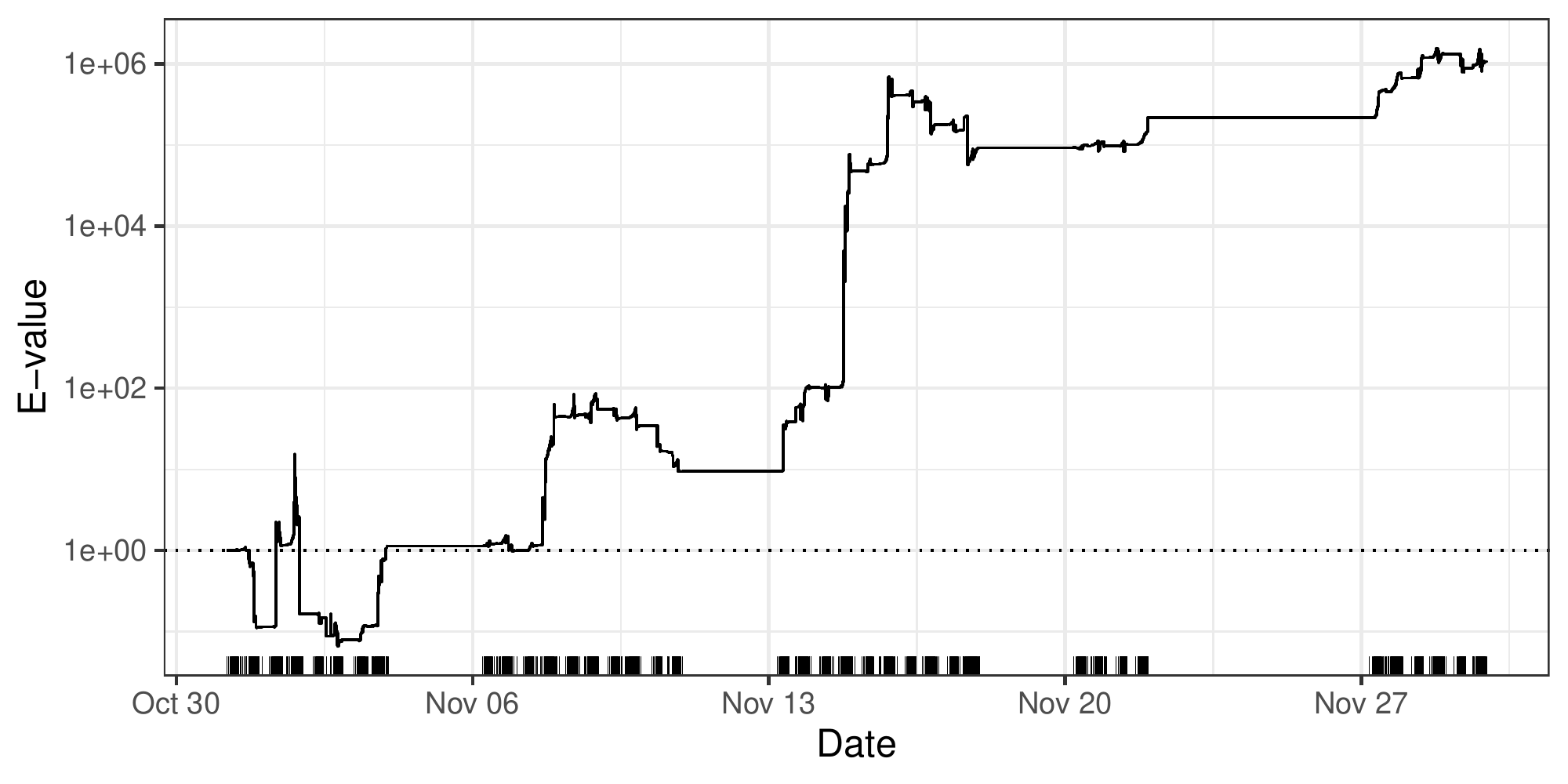}
		\caption{\textit{E}-value for the bike sharing data set. Lines at the bottom indicate at which times observations occurred. \label{fig:data_application}}	    
	\end{center}
	\end{figure}

	\citet{Berrett2020} apply the conditional randomization test with the test statistic $|\mathrm{cor}(Y, X - \mathbb{E}_{Q_Z}[X])|$. Applied to the November data, this yields a p-value of essentially zero: the observed value of the test statistic was greater than any value obtained with simulated $X$, over $10\,000$ simulations. For the \E-statistics, we model the probability that $Y = 1$ with logistic regression, taking $X$ and $\mu_{Z}$ as covariates and starting with a minimal sample size of $200$ observations in the test data. A full model including $Z$ instead of only $\mu_{Z}$ would be problematic due to the high number of combinations of starting points and destinations relative to the size of the test data. We do not have this limitation in the estimation of the distribution of $X\mid Z$ due to the much larger size of the training data set. However, for a valid test it is not necessary to include $Z$ itself in the model. The probability predictions from the logistic model are then truncated to the interval $[0.01, 0.99]$. For the sequential analysis, the rides are arranged in the order of the start date and time of ride. Figure \ref{fig:data_application} shows how evidence accumulates over time. At the end of the period, an \E-value of more than $10^6$ is attained, giving decisive evidence against the null hypothesis of conditional independence. An \E-value of $10^4$ is already reached at the $4380$th observation, on November 15th, and hence after about half of the total observation time.

	\section{Discussion, Related and Future Work}\label{sec:discuss}
	We have proposed and analyzed anytime-valid tests of conditional independence in the model-X setting. 
	Our method gives a general procedure to transform statistics that measure conditional independence to \E-statistics.
	We have shown that for a simple alternative, using the conditional density as statistic leads to the growth rate optimal \E-statistic, and derived a bound on the inflation of Type-I error under violations of the MX assumption.
	
	\citet{Duan2022} and \citet{Shaer2022} have also proposed methods to test CI under MX, but they address the problem in fundamentally different ways than we do. 
    In the fully sequential setting, \citet[Appendix E.4]{Duan2022} construct a test martingale which grows if a researcher can better than randomly predict a binary $X_n$ given past information and $(Y_n, Z_n)$; notice, however, that this is just a small part of their work, which, for example, also includes tests for a batch setting.
	\citet{Shaer2022} construct a test martingale which measures how much better a forecaster can predict $Y_n$ based on past data and general $(X_n, Z_n)$, relative to a forecaster only having access to a randomized $\tilde{X}_n \sim Q_{Z_n}$. 
	A derandomization of this procedure is obtained by taking the expectation over $\tilde{X}_n$, which yields similar test martingales as in Theorem \ref{thm:e_var}.
	Both of these methods are very flexible in the choice and tuning of the prediction method, and may therefore behave quite differently from ours. 
	However, ours are justified in terms of the strong GRO optimality criterion by Proposition \ref{prop:misspec} and Corollary~\ref{cor:asympgro}, whenever a fast-converging estimator is used. 
	While this makes our approach in some sense the optimal one, it requires specifying a reasonable (potentially nonparametric) set of densities ${\cal F}$ to define the estimator $\hat{f}$.
    We can basically use any set of densities  we like, as long as the estimators converge in information in the sense below Corollary~\ref{cor:asympgro}. However, it may not always be easy to find ${\cal F}$ for which estimation is computationally feasible and practically successful, especially if $Z$ is high-dimensional. In that case the approaches of \citet{Duan2022} and \citet{Shaer2022} may have the advantage of being more flexible
    --- whether or not this is the case may be domain-dependent, and is an interesting avenue for future research. 
	Finally, the issue of robustness with respect to misspecification of the distribution of $X$ given $Z$ is not addressed by \citet{Duan2022} and only assessed by \citet{Shaer2022} in a limited number of settings through simulations, whereas our Theorem \ref{thm:bound} yields the same worst-case bound for our approach as in the batch setting.
	
    For a comparison of our method to the CRT, the following aspects are worth highlighting.
    Our simulations suggest that our anytime-valid tests with optional stopping do not need more samples, on average, to achieve the same power as the CRT. 
    To be precise, for a given desired power, researchers have to plan for a higher maximum sample size with our anytime-valid tests. 
    But thanks to early stopping, the average sample size of experiments is not more or even less than for the classical CRT. 
    This confirms the findings by \citet{Gruenwald2019,terSchure2021, Turner2021} on other anytime-valid tests and their nonsequential counterparts. 
    In terms of computational complexity, our method is not less efficient than the classical CRT if the functions $h_n$ in the test martingale \eqref{eq:test_martingale} can be updated in constant time; in our application on logistic regression we did not make use of recursive updating, though. 
    A slight advantage of our method compared to the fixed sample size CRT is that the latter requires at least $\lceil 1/\alpha-1\rceil$ resamples in order to be able to obtain a p-value below $\alpha$, whereas the test martingale $(S^{\CI}_{h^n})_{n\in\mathbb{N}}$ can exceed any level independently of the number $M$ of resamples. 
    If the strategy of Proposition \ref{prop:integral} is applied, then each factor in $S^{\CI}_{h^n}$ is bounded by $M+1$, but not their cumulative product.
    
	There are various avenues for future research. With respect to robustness, our Theorem \ref{thm:bound} shows that for a fixed upper bound $N$ on the sample size, the sequential test achieves the same worst case inflation of rejection rate as the nonsequential CRT with a sample size $N$. It would be of interest to investigate the robustness of the test when this upper bound grows and the approximation $\hat{Q}_Z$ of the conditional distributions of $X\mid Z$ are sequentially updated with new samples. Furthermore, our simulations indicate that our \E-statistics have competitive power compared to existing methods with relatively small dimension of $Z$, but as stated above, further research is necessary to investigate suitable \E-statistics and their power when $Z$ is of higher dimension.
	
	Finally, the \E-statistics in this paper are highly tailored to the MX assumption, and it is an open question to us how to construct general sequential tests of conditional independence without the MX assumption. 
	However, the construction strategy hints at a more general scheme for creating \E-statistics which might be useful both in conditional independence and other testing problems. 
	Namely, suppose that general observations $D$ in some space $\mathcal{D}$ are available, and we are interested in testing a null hypothesis $\mathcal{H}_0$ about the distribution of $D$. If for two functions $g, h$ with $h > 0$, the distribution of $h(D)$ given $g(D)$ is known to equal $Q_{g(D)}$ under all elements of $\mathcal{H}_0$, then
	$
	    E(D) = h(D)/\int h(t) \, \mathrm{d}Q_{g(D)}(t)
	$
	is an \E-statistic for $\mathcal{H}_0$. The conditional independence \E-statistics are of this form with $D = (X,Y,Z)$, arbitrary $h$, and $g(D) = (Y,Z)$. If $\mathcal{D} = \mathbb{R}$ and $\mathcal{H}_0$ contains all distributions which are symmetric around zero, one obtains an \E-statistic for any $h$ by choosing $g(D)=|D|$,
	\[
	    E(D) = \frac{h(D)}{\int h(t) \, \mathrm{d}Q_{g(D)}(t)} = \frac{h(D)}{(h(D) + h(-D))/2}.
	\]
	Choosing $h(D) = \exp(\lambda D)$ for $\lambda \in \mathbb{R}$ yields the so-called Efron-De la Pe\'na \E-statistic, first described by \citet{DeLaPena1999} and studied in recent years by \citet{Ramdas2020,Kant2021EvaluatingThesis}. This strategy may be fruitful for constructing tests of other hypotheses.

    \if1\blind
    {
	\section*{Acknowledgments}
	Alexander Henzi gratefully acknowledges support from the Swiss National Science Foundation. Computations have been performed on UBELIX (\url{https://ubelix.unibe.ch/}), the HPC cluster of the University of Bern. We are grateful to Aaditya Ramdas, Yaniv Romano and three anonymous referees for their helpful feedback on this article.
	}
	
	\bibliographystyle{JASA/agsm}
	\bibliography{logistic_regression_e_values_biblio}
	
	\appendix
	
	\section{Additional Simulations} \label{app:extra_sim}
	\subsection{Effect of Truncation on Power}
\begin{figure}[ht]
	\begin{center}
	    \includegraphics[width = .75\textwidth]{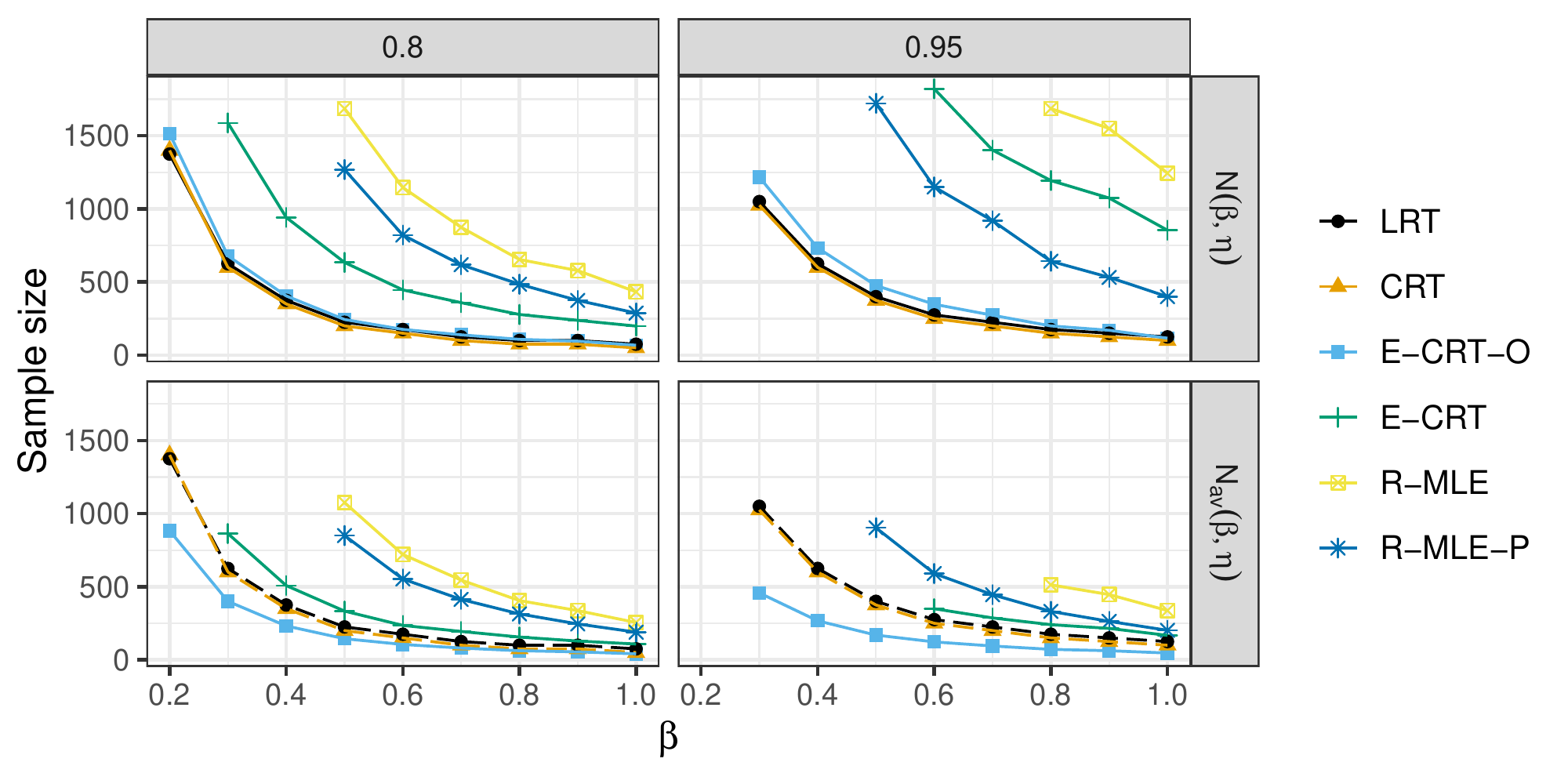}
	    \caption{Sample sizes for different methods as in Figure \ref{fig:alternative} but with $\varepsilon = 0$ for the \E-statistics.}
	    \label{fig:alternative_eps0}
	\end{center}
	\end{figure}

	Figure \ref{fig:alternative_eps0} shows the same plot as Figure \ref{fig:alternative} but without truncation for the probabilities in the \E-statistics, i.e.~$\varepsilon = 0$. 	
	
	\subsection{Robustness With Respect to Misspecification}\label{sec:robustness}
	We test the robustness of the randomization based \E-statistics with respect to misspecification of the conditional distribution of $X$ in the same way as in the simulation study of \citet{Berrett2020}. 
	All simulations in this section are under the null hypothesis, i.e.~$\beta = 0$. Rejection rates of the \E-statistics are again computed with a maximal sample size of $2000$ and with optional stopping, i.e.~rejection if the level $1/\alpha$ is exceeded at least once, and with truncation level $\varepsilon = 0.05$. For comparison, the conditional randomization test is applied to a sample of fixed size, for sizes $200$, $1000$, and $2000$, and additionally with the unconditional absolute correlation $|\mathrm{cor}(X,Y)|$ as test statistic, as in \citet{Berrett2020}, for sample sizes $200$ and $2000$.
	
\begin{figure}[ht]
	    \begin{center}
	    \includegraphics[width=.8\textwidth]{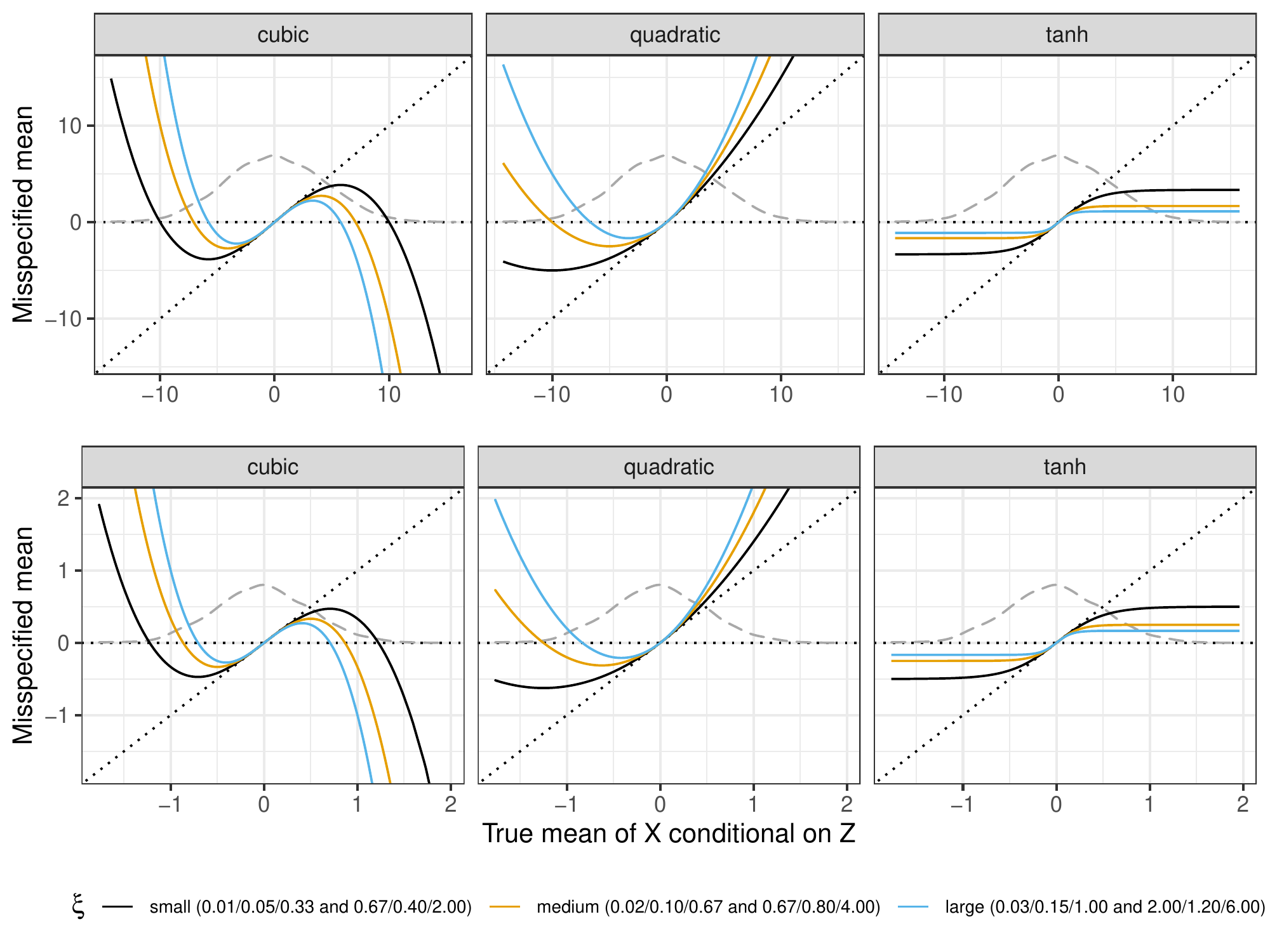}
		\caption{Misspecification in the conditional mean of $X$ given $Z$ for the three different functions from Section \ref{sec:robustness}. Upper row of plots: $X\mid Z$ generated as in \citet[Section 6.1.1]{Berrett2020}. Lower row: $X\mid Z$ generated as in Section \ref{sec:sim} with $q = 4$. The dashed line shows the (height adjusted) density of the conditional expectation of $X$ given $Z$. The values for $\xi$ given in the legend refer to the misspecifications in the same order as the panel colums (cubic/quadratic/tanh), with the first triple giving $\xi$ for $X\mid Z$ as simulated by \citet{Berrett2020} (upper three figures), and the second triple the values of $\xi$ applied when $X\mid Z$ is generated as in Section \ref{sec:sim} (lower three figures). \label{fig:misspecification_function}}    
	    \end{center}
	\end{figure}

\begin{figure}[ht]
	\begin{center}
	    \includegraphics[width=.8\textwidth]{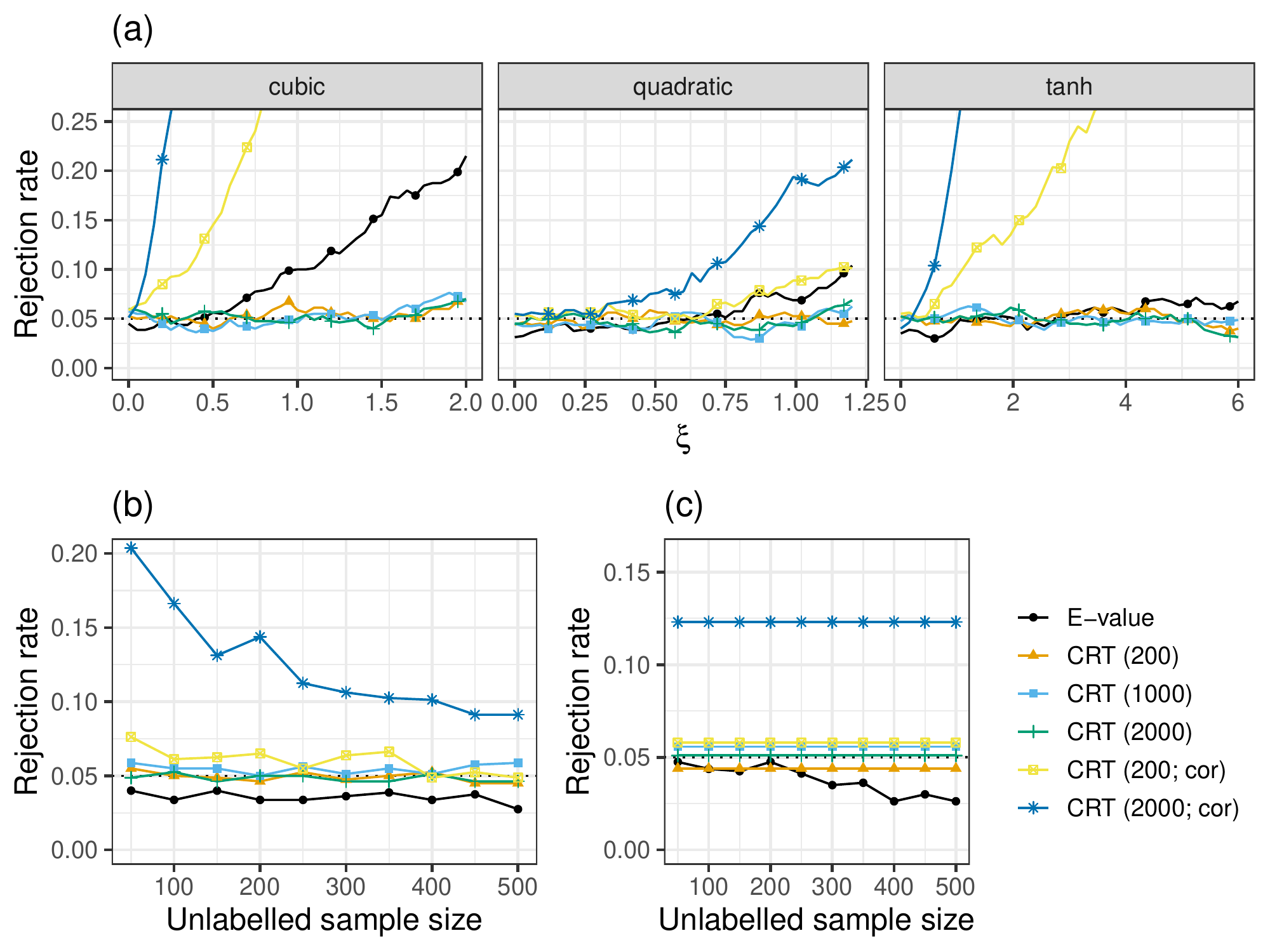}
		\caption{(a) Rejection rates of \E-values and conditional randomization test (with sample sizes $n = 200, \, 1000, \, 2000$ and likelihood as test statistic, and $n = 200, \, 2000$ and correlation as test statistic) at the level $\alpha = 0.05$, under different misspecifications for the conditional mean of $X$. (b) Rejection rates when the distribution of $X$ is estimated on a separate sample, for varying sample sizes. (c) Rejection rates when the same data is used both for estimating the conditional distribution of $X$ and applying the test, as described in the text. \label{fig:robustness}}
	\end{center}
	\end{figure}

	First, instead of sampling $X$ with conditional mean $\mu_Z$ as defined in \eqref{eq:true_conditional_mean}, we set the mean to
	\begin{align*}
		& \mu_Z - \xi \mu_Z^3 & \ \text{(cubic misspecification)}, \\
		& \mu_Z + \xi \mu_Z^2 & \ \text{(quadratic misspecification)}, \\
		& \mathrm{tanh}(\xi\mu_Z)/\xi & \ \text{(hyperbolic tangent)},
	\end{align*}
	which are the same misspecifications as in \citet[Section 6.1.1]{Berrett2020}. 
	They are illustrated in Figure~\ref{fig:misspecification_function} for different values of $\xi$, the range of which has been selected for each misspecification type in such a way that the relative misspecification compared to the true mean approximately matches the one in the simulations by \citet{Berrett2020}.
	When the parameter $\xi$ equals zero, understood as limit $\xi \rightarrow0$ for the hyperbolic tangent, the model is correctly specified. 
	Panel (a) of Figure \ref{fig:robustness} shows that both the CRT and the \E-statistics are robust with respect to slight misspecifications of the conditional mean. 
	The CRT based on the likelihood is much more robust than the other two tests, due to the fact that re-estimating the logistic regression model with simulated $X$ is invariant under affine transformations of $X$ and $Z$ and hence able to correct much of the misspecification. The \E-statistic based test is less robust than this variant of the CRT, since it does not re-estimate the logistic model with simulated $X$, but still substantially more robust than the CRT based on unconditional correlation, which already with $n = 200$, as compared to $n = 2000$ for the \E-values, has rejection rates strongly exceeding the nominal level as $\xi$ increases.
	
	In panel (b) of Figure \ref{fig:robustness}, the rejection rates of the tests are shown when the distribution of $X_p$ is estimated on an independent unlabeled data set, for different sizes of this data set. The estimation of the conditional distribution is by linear regression, with the maximum likelihood estimator for the conditional variance. Here the \E-statistics have rejection rates below the nominal level, even for unlabeled sample size as small as $50$. Also the CRT with logistic likelihood as test statistic has rejection rate close to the nominal level.
	
	Finally, in panel (c) of Figure \ref{fig:robustness} the rejection rates are depicted for the case when the same data is used both for estimating the distribution of $X$ and for testing. The estimation is as described in the previous paragraph. For the CRT, the distribution of $X$ is estimated on the same data to which the test is applied, like in the simulation study of \citet{Berrett2020}. For the \E-statistics, a slightly different approach is taken, tailored to sequential settings. We start with a potentially small unlabeled sample, and each time a new instance is observed, the estimate of the distribution of $X$ is updated with all the data available so far. Again, all tests except for the correlation based CRT with sample size $2000$ have rejection rate close to the nominal level.

	\section{Proofs of Main Results}\label{app:proofs}
	
	\subsection{Proof of Theorem~\ref{thm:e_var}} \label{app:proof_e_var}
	\begin{proof}
		Let $P\in \mathcal{H}_0$ arbitrarily.
		The proof relies on the simple insight that we can separate the expectation with respect to $(Y_n,Z_n)$ from that with $X_n$,
		\begin{align*}
			\mathbb{E}_{P}[E_{h_n}^{\CI}(X_n,Y_n,Z_n)\mid D^{n-1}] & = \mathbb{E}_{P}\left[\mathbb{E}_{P}\middle[\frac{h_n(X_n, Y_n , Z_n)}{\int_{\mathcal{X}} h_n(x, Y_n, Z_n) \, \mathrm{d}Q_{Z_n}(x)}\middle | Y_n, Z_n, D^{n-1}\middle] \, \middle| \, D^{n-1} \right] \\
			& = \mathbb{E}_P \left[ \frac{\int_{\mathcal{X}} h_n(x', Y_n, Z_n) \, \mathrm{d}Q_{Z_n}(x')}{\int_{\mathcal{X}} h_n(x, Y_n, Z_n) \, \mathrm{d}Q_{Z_n}(x)} \middle | D^{n-1} \right]
			= 1,
		\end{align*}
		where in the last step we use that $P_{X_n|Y_n,Z_n} = P_{X_n|Z_n} = Q_{Z_n}$.
	\end{proof}
	
	\subsection{Proof of Proposition \ref{prop:integral}}
	
	\begin{proof}
	Define $\tilde{X}_0 = X_n$. The random variables $\tilde{X}_0, \dots, \tilde{X}_M$ are exchangeable, so
	\[
	    \check{E}_{h_n;j}^{\CI}(D_n) := \frac{h_n(\tilde{X}_j, Y_n , Z_n)}{\sum_{i=0}^M h_n(\tilde{X}_i, Y_n, Z_n)/(M+1)}, \ j = 0, \dots, M,
	\]
	have the same expected value as $\check{E}_{h_n}^{\CI}(D_n) = \check{E}_{h_n;0}^{\CI}(D_n)$. Since $\sum_{i=0}^M \check{E}_{h_n;i}^{\CI}(D_n) \equiv M+1$, this implies $\mathbb{E}_P[\check{E}_{h_n}^{\CI}(D_n)\mid D^{n-1}] = 1$.
	\end{proof}

	\subsection{Proof of Theorem~\ref{thm:GRO}} \label{app:proof_GRO}
	
	\begin{proof}
		Let $f = f_{X,Y,Z}(x,y,z)$ be the density of $(X,Y,Z)$ with respect to a measure $\sigma\times\mu\times\nu$ on $\mathcal{X}\times\mathcal{Y}\times\mathcal{Z}$. Then the conditional density $f_{Y\mid X,Z}$ equals
		\[
		f_{Y\mid X, Z}(y\mid x, z) = \frac{f(x, y, z)}{\int_{\mathcal{Y}} f(x, s, z) \, \mathrm{d}\sigma(s)}.
		\]
		The density of $Q_Z$ must equal the conditional density $f_{X\mid Z}$, which is given by
		\[
		f_{X \mid Z}(x\mid z) = \frac{\int_{\mathcal{Y}}f(x, s, z) \, \mathrm{d}\sigma(s)}{\int_{\mathcal{X}} \int_{\mathcal{Y}} f(r, s, z) \, \mathrm{d}\sigma(s) \, \mathrm{d}\mu(r)},
		\]
		so that, with $h(x,y,z) = f_{Y|X,Z}(y|x,z)$,
		\[
		\int_{\mathcal{X}} h(x,y,z) \, \mathrm{d}Q_z(x) = \int_{\mathcal{X}} \frac{f(r, y, z)}{\int_{\mathcal{X}} \int_{\mathcal{Y}} f(r', s, z) \, \mathrm{d}\sigma(s) \, \mathrm{d} \mu(r')} \,\mathrm{d}\mu(r) = f_{Y \mid Z}(y\mid z).
		\]
		Hence the \E-statistic with this choice of $h$ is equal to
		\[
		E^{\CI}_{f_{Y\mid X,Z}}(X_i,Y_i,Z_i) = \frac{f_{Y\mid X,Z}(Y_i \mid X_i, Z_i)}{f_{Y \mid Z}(Y_i\mid Z_i)} \\
		= \frac{f_{X, Y, Z}(X_i, Y_i, Z_i)}{f_{Y \mid Z}(Y_i\mid Z_i) f_{X \mid Z}(X_i\mid Z_i)f_{Z}(Z_i)}.
		\]
		The denominator is the density of an element of $\mathcal{H}_0$ as in~\eqref{eq:null_hypothesis}.
		Theorem 1 by \citet{Gruenwald2019} states that this \E-statistic must therefore be the GRO \E-statistic for a single data point $(X_i, Y_i, Z_i)$ and 
		the same argument can be applied to the product of these \E-statistics.
		Finally, a slight rewriting shows that the \E-statistic corresponds to the ratio of the joint conditional density of $(X, Y)$ given $Z$ divided by the product of its marginals. 
		For all $i$, the expected value of $\log E^{\CI}_{f_{Y\mid X,Z}}(X_i,Y_i,Z_i)$ conditional on $Z$ is therefore equal to the conditional mutual information of $X$ and  $Y$ given $Z$.
	\end{proof}
	
	\subsection{Proof of Proposition~\ref{prop:misspec}}
	\begin{proof}
		Since the distribution $Q_Z$ is well-specified, we denote $g_{Y\mid Z}$ for the density $\int g_{Y\mid X,Z} \, \mathrm{d}Q_Z$. 
		Then the quantity of interest is given by
		\begin{align*}
			\mathbb{E}_f&\left[\log E^{\CI}_{g_{Y\mid X,Z}}(x,y,z)\right]=\mathbb{E}_f\left[\log \frac{g_{Y\mid X, Z}(y \mid x, z)}{g_{Y \mid Z}(y\mid z)}\right]\\
			&=I_f(X;Y\mid Z) +\mathbb{E}_f\left[\log \frac{g_{Y\mid X, Z}(y \mid x, z)}{g_{Y \mid Z}(y\mid z)}-\log \frac{f(x, y, z)}{f_{X\mid Z}(x\mid z)f_{Y \mid Z}(y\mid z)f_Z(z)} \right]\\
			&=I_f(X;Y\mid Z) +\mathbb{E}_f\left[\log \frac{g_{Y\mid X, Z}(y \mid x, z)}{g_{Y \mid Z}(y\mid z)}-\log \frac{f_{Y\mid X,Z}(y\mid x, z)}{f_{Y \mid Z}(y\mid z)} \right]\\
			&=I_f(X;Y\mid Z)+\mathbb{E}_f[\mathrm{KL}(f_{Y\mid Z}\| g_{Y\mid Z})]-\mathbb{E}_f[\mathrm{KL}(f_{Y\mid X,Z}\| g_{Y\mid X,Z})].
		\end{align*}
		The desired result follows from the nonnegativity of KL divergence.
	\end{proof}
	
	\subsection{Proof of Theorem~\ref{thm:bound}}
	\begin{proof}
		Fix $N \in \mathbb{N}$ and $\alpha \in (0,1)$. Conditional on $Y_i, Z_i$, $i = 1, \dots, N$, the randomness of the process $S_n = S_n(X^n) = \prod_{i=1}^n \tilde{E}^{\CI}_{h_n}$, $n = 1, \dots, N$, solely stems from $X_1, \dots, X_N$, and we will write $Y_i$, $Z_i$ with lower case letters $y_i$, $z_i$ to reflect that all statements are conditional on their values. So the \E-value at time $n$ writes as
		\[
		\tilde{E}^{\CI}_{h_n} = \frac{h_n(X_n, y_n, z_n \mid X^{n-1}, y^{n-1}, z^{n-1})}{\int_{\mathcal{X}}h_n(x, y_n, z_n \mid X^{n-1}, y^{n-1}, z^{n-1}) \, \mathrm{d}\hat{Q}_{z_n}(x)}.
		\]
		The condition $h_n > 0$ ensures that this \E-value is well-defined. For $n > N$, set $h_n \equiv 1$, so that $S_n = S_N$ for $n > N$. If $X^N$ has distribution $\hat{Q}^N_{Z^N}$, then the process $(S_n)_{n \in \mathbb{N}}$ is a nonnegative martingale with respect to the filtration $\mathcal{F}_n = \sigma(X_1, \dots, X_n)$, because
		\begin{align*}
			\mathbb{E}\left[ S_n \vert X^{n-1}\right] =
			\quad \int_{\mathcal{X}}  \frac{h_n(x, y_n, z_n \mid X^{n-1}, y^{n-1}, z^{n-1})}{\int_{\mathcal{X}}h_n(x, y_n, z_n \mid X^{n-1}, y^{n-1}, z^{n-1}) \, \mathrm{d}\hat{Q}_{Z_n}(x)} \, \mathrm{d}\hat{Q}_{Z_n}(x) = 1
		\end{align*}
		almost surely. Hence by Ville's inequality, $P\left(\exists \, n \leq N \colon S_n \geq 1/\alpha\right) \leq \alpha$.
		Let
		\[
			A = \left\{x^n \in \mathcal{X}^n\colon \exists \, n \leq N \text{ s.t. } S_n(x^n) \geq 1/\alpha\right\}.
		\]
		Then, since $Q^N_{Z^N}(A) = P(\exists \, n \leq N\colon S_n \geq 1/\alpha \mid Y^N = y^N, Z^N = z^N)$,
		\[
		P(\exists \, n \leq N\colon S_n \geq 1/\alpha \mid Y^N, Z^N) \leq \hat{Q}^N_{z_N}(A) + d_{\mathrm{TV}}(Q^N_{Z^N}, \hat{Q}^N_{Z^N}) \leq \alpha + d_{\mathrm{TV}}(Q^N_{Z^N}, \hat{Q}^N_{Z^N}).
		\]
	\end{proof}
	
	\subsection{Proof of Proposition~\ref{prop:logistic}} \label{app:proof_logistic}
	
	\begin{proof}
		The subgaussianity assumption (i) implies that 
		\begin{equation} \label{eq:covariate_tailprob}
			P\left(|u^{\top}((X,Z) - \mathbb{E}[(X,Z)])| \geq \eta) \leq 2\exp(-\eta^2/(2\|u\|^2\sigma^2)\right), \ \eta > 0, \ u \in \mathbb{R}^{p+q},
		\end{equation}
		and that $\mathbb{E}[\|(X,Z)\|^k] < \infty$ for all $k \in \mathbb{N}$. 
		As a consequence of the latter and of assumption (i)(a), Theorem 1 of \cite{Qian2002} implies that the MLE $\hat{\theta}_n$ exists with asymptotic probability one and satisfies $\|\hat{\theta}_n - \theta\| = \mathcal{O}( n^{-1/2}\log(\log(n))^{1/2})$ almost surely.
		
		We now study the properties of the function $\theta \mapsto \log(p_{\theta}(y \mid x, z))$ for $\theta \in \mathbb{R}^{p+q}$.
		The derivative of $\log(p_{\theta}(y \mid x, z))$ with respect to $\theta_j$ equals
		\[
		\frac{d}{d\theta_j} \log(p_{\theta}(y \mid x, z)) = \begin{cases}
		yx_{j} - x_{j} p_{\theta}(1 \mid x, z) & \text{if}\, j\leq p\\  yz_{j-p}, - z_{j-p} p_{\theta}(1 \mid x, z) & \text{else}.\end{cases}
		\]
		Consequently, for any $\theta, \theta' \in \mathbb{R}^{p+q}$,
		\begin{equation} \label{eq:lipschitz}
			|\log(p_{\theta}(y \mid x, z)) - \log(p_{\theta'}(y \mid x, z))| \leq \|(x,z)\|\|\theta - \theta'\|.
		\end{equation}
		This implies that
		\begin{align*}
			\frac{1}{n}\Big\vert\sum_{i=1}^n\log(p_{\hat{\theta}_{i-1}}(Y_i \mid X_i, Z_i)) - \log(p_{\theta}(Y_i \mid X_i, Z_i)) \Big\vert & \leq \frac{1}{n}\sum_{i=1}^n \|\hat{\theta}_{i-1} - \theta\|\|(X_i,Z_i)\| \\
			& \leq \left(\frac{1}{n}\sum_{i=1}^n \|\hat{\theta}_{i-1} - \theta\|^2\right)^{1/2}\left(\frac{1}{n}\sum_{i=1}^n \|(X_i,Z_i)\|^2\right)^{1/2}.
		\end{align*}
		Since $\|(X_i,Z_i)\|^2$, $i \in \mathbb{N}$, are independent with expectation $\mathbb{E}[\|(X,Z)\|^2] < \infty$, the law of large number implies that $\sum_{i=1}^n \|(X_i,Z_i)\|^2/n \rightarrow \mathbb{E}[\|(X,Z)\|^2] < \infty$ almost surely, and $\sum_{i=1}^n \|\hat{\theta}_{i-1} - \theta\|^2/n \rightarrow 0$ since $\|\hat{\theta}_{n} - \theta\| \rightarrow 0$ almost surely as $n \rightarrow \infty$. It remains to show an analogous convergence result for the denominator in $S^{CI}_n$. Define
		\[
		r_n = \frac{\int p_{\theta}(Y_n \mid x, Z_n) \, \mathrm{d}Q_{Z_n}(x)}{\int p_{\hat{\theta}_{n-1}}(Y_n \mid x, Z_n) \, \mathrm{d}Q_{Z_n}(x)}.
		\]
		We want to show that $\liminf_{n\rightarrowtail\infty}\sum_{i=1}^n \log(r_i)/n \geq 0$ almost surely. To this end, write
		\begin{align*}
			r_n & = \frac{\int p_{\theta}(Y_n \mid x, Z_n) \, \mathrm{d}Q_{Z_n}(x)}{\int p_{\theta}(Y_n \mid x, Z_n) \, \mathrm{d}Q_{Z_n}(x) + \int (p_{\hat{\theta}_{n-1}}(Y_n \mid x, Z_n) - p_{\theta}(Y_n \mid x, Z_n)) \, \mathrm{d}Q_{Z_n}(x)} \\
			& \geq \frac{\int p_{\theta}(Y_n \mid x, Z_n) \, \mathrm{d}Q_{Z_n}(x)}{\int p_{\theta}(Y_n \mid x, Z_n) \, \mathrm{d}Q_{Z_n}(x) + \int |p_{\hat{\theta}_{n-1}}(Y_n \mid x, Z_n) - p_{\theta}(Y_n \mid x, Z_n)| \, \mathrm{d}Q_{Z_n}(x)}.
		\end{align*}
		Since $\log(1+x) \leq x$, we have $\log(1/(1+x)) = -\log(1+x) \geq -x$, for $x > -1$. So
		\[
		\log(r_n) \geq -\frac{\int |p_{\hat{\theta}_{n-1}}(Y_n \mid x, Z_n) - \int p_{\theta}(Y_n \mid x, Z_n)| \, \mathrm{d}Q_{Z_n}(x)}{\int p_{\theta}(Y_n \mid x, Z_n) \, \mathrm{d}Q_{Z_n}(x)}
		\]
		The function $\theta \mapsto p_{\theta}(y \mid x, z)$ is Lipschitz continuous, because for $k = 1, \dots, p + q$,
		\[
		\Big\vert\frac{d}{d\theta_k} p_{\theta}(y \mid x, z) \Big\vert =
		\begin{cases}
		    |x_k|p_{\theta}(1 \mid x, z)(1-p_{\theta}(1 \mid x, z)) \leq |x_k| & \text{if } k = 1, \dots, p \\
		    |z_{k-p}|p_{\theta}(1 \mid x, z)(1-p_{\theta}(1 \mid x, z)) \leq |z_{k-p}| & \text{else}. \\
		\end{cases}
		\]
		This implies that
		\[
		\log(r_n) \geq -\frac{\|\hat{\theta}_{n-1} - \theta\| \int \|(x, Z_n)\| \, \mathrm{d}Q_{Z_n}(x)}{\int p_{\theta}(Y_n \mid x, Z_n) \, \mathrm{d}Q_{Z_n}(x)}.
		\]
		To bound this from below, we now show that the denominator $\int p_{\theta}(Y_n \mid x, Z_n) \, \mathrm{d}Q_{Z_n}(x)$ is small only with a small probability. Let $\kappa_n = n^{-\delta}/2$ for $\delta > 0$. Define the events
		\[
		A_n = \left\{\min_{y=0,1}p_{\theta}(y \mid X_n, Z_n) \leq \kappa_n\right\}.
		\]
		Let $\mathrm{logit}(p) = \log(p/(1-p))$. Then,
		\[
		\min_{y = 0,1}p_{\theta}(y \mid x, z) \leq \kappa_n  \iff |\theta^{\top}(x,z)| \geq |\mathrm{logit}(\kappa_n)|,
		\]
		and therefore, since $|\mathrm{logit}(p)| \geq |\log(2p)|$ for $p \in (0,1/2]$,
		\[
		A_n \subseteq \{|\theta^{\top}(X_n,Z_n)| \geq |\log(2\kappa_n)|\} = \{|\theta^{\top}(X_n,Z_n)| \geq \delta\log(n)\},
		\]
		The above derivations yield $P(A_n) \leq P(|\theta^{\top}(X_n,Z_n)| \geq\delta\log(n))$, and \eqref{eq:covariate_tailprob} implies, with $B = \|\theta\|$,
		\begin{align*}
		    P(|\theta^{\top}(X,Z)| \geq\delta\log(n)) &  \leq P\left(|\theta^{\top}((X,Z) - \mathbb{E}[(X,Z)])| \geq\delta\log(n)) - |\theta^{\top}\mathbb{E}[(X,Z)]|\right) \\ 
		    & \leq 2\exp(-\delta^2\log(n)^2/(8B^2\sigma^2)),
		\end{align*}
		for $n$ large enough such that $\delta\log(n)/2 \geq |\theta^{\top}\mathbb{E}[(X,Z)]|$.
		In a next step, we use this to bound $\min_{y=0,1}\int p_{\theta}(y \mid x, Z_n) \, \mathrm{d}Q_{Z_n}(x)$. 
		First, note that for $y\in \{0,1\}$,
		\begin{align*}
		    \int p_{\theta}(y \mid x, Z_n) \, \mathrm{d}Q_{Z_n}(x) & = \int p_{\theta}(y \mid x, Z_n)1\{p_{\theta}(y \mid x, Z_n) \geq 1-\kappa_n\} \, \mathrm{d}Q_{Z_n}(x) \\
			& \quad + \int p_{\theta}(y \mid x, Z_n)1\{p_{\theta}(y \mid x, Z_n) < 1-\kappa_n\} \, \mathrm{d}Q_{Z_n}(x) \\
			& \leq Q_{Z_n}(p_{\theta}(y \mid X_n, Z_n) \geq 1-\kappa_n) + 1-\kappa_n.
		\end{align*}
		It follows that for $\eta>0$, if $\int p_{\theta}(y \mid x, Z_n) \, \mathrm{d}Q_{Z_n}(x) \geq 1-n^{-\eta}$, then $Q_{Z_n}(p_{\theta}(y \mid X_n, Z_n)\geq 1-\kappa_n) \geq \kappa_n - n^{-\eta}$.
        Recall that $\kappa_n = n^{-\delta}/2$ with $\delta > 0$ unspecified so far. 
        For $n$ large enough such that $n^{-\eta/2}\leq 1/4$, choosing $\delta = \eta/2$ implies $\kappa_n-n^{-\eta}=n^{-\eta/2}(1/2 - n^{-\eta/2})\geq n^{-\eta/2}/4$.
		Consequently, for large $n$, by Markov's inequality,
		\begin{align}
			P\left(\int p_{\theta}(y \mid x, Z_n) \, \mathrm{d}Q_{Z_n}(x) \geq 1-n^{-\eta}\right) & \leq P \left(Q_{Z_n}(p_{\theta}(y \mid X_n, Z_n)\geq 1-\kappa_n) \geq n^{-\eta/2}/4\right) \nonumber \\
			& \leq 4n^{\eta/2} \mathbb{E}[Q_{Z_n}(p_{\theta}(y \mid X_n, Z_n)\geq 1-\kappa_n)] \nonumber \\
			& = 4n^{\eta/2} P(p_{\theta}(y \mid X_n, Z_n)\geq 1-\kappa_n). \label{eq:small_prob}
		\end{align} 
		But it has already been shown that
		\[
		P(p_{\theta}(y \mid X_n, Z_n)\geq 1-\kappa_n) = P(p_{\theta}(1-y \mid X_n, Z_n)\leq \kappa_n) \leq 2\exp(-\delta^2\log(n)^2/(8B^2\sigma^2))
		\]
		for large $n$, which in \eqref{eq:small_prob} gives an upper bound of 
		\[
		8\exp\left(-(\eta/2)^2\log(n)^2/(8B^2\sigma^2)+\log(n)\eta/2\right) = 8\exp\left(-\log(n)(\eta^2\log(n)/(32B^2\sigma^2) - \eta/2)\right).
		\]
		Since $\eta^2\log(n)/(32B^2\sigma^2) - \eta/2 \rightarrow \infty$ as $n \rightarrow \infty$, it holds that $\eta^2\log(n)/(32B^2\sigma^2) - \eta/2 > 1$ for $n$ large enough, and we can conclude
		\[
		\sum_{n=1}^\infty P\left(\min_{y=0,1} \int p_{\theta}(y \mid x, Z_n) \, \mathrm{d}Q_{Z_n}(x) \leq n^{-\eta}\right) < \infty.
		\]
		Thus the Borel-Cantelli Lemma implies that $\min_{y=0,1}\int p_{\theta}(y \mid x, Z_n) \, \mathrm{d}Q_{Z_n}(x) \leq n^{-\eta}$ holds for only finitely many $n$ with probability one. Now
		\begin{align}
			\frac{1}{n}\sum_{i=1}^n \log(r_i) & \geq -\frac{1}{n} \sum_{i=1}^{n} \frac{\|\hat{\theta}_{i-1} - \theta\| \int \|(x, Z_i)\| \, \mathrm{d}Q_{Z_i}(x)}{\int p_{\theta}(Y_i \mid x, Z_i) \, \mathrm{d}Q_{Z_i}(x)} \nonumber \\
			& \geq -\frac{M}{n} -\sum_{i=1}^{n} i^{\eta} \|\hat{\theta}_{i-1} - \theta\| \int \|(x, Z_i)\| \, \mathrm{d}Q_{Z_i}(x) \nonumber \\
			& = -\frac{M}{n} -\frac{1}{n}\sum_{i=1}^{n} i^{\eta} \|\hat{\theta}_{i-1} - \theta\| \mathbb{E}[\|(X_i,Z_i)\| \mid Z_i] \nonumber \\
			& \geq -\frac{M}{n} -\left(\frac{1}{n}\sum_{i=1}^{n} i^{2\eta} \|\hat{\theta}_{i-1} - \theta\|^2\right)^{1/2} \left(\frac{1}{n}\sum_{i=1}^{n} \mathbb{E}[\|(X_i,Z_i)\| \mid Z_i]^2\right)^{1/2}, \label{eq:lower_bound_r}
		\end{align} 
		where
		\[
		M = \sum_{i=1}^{\infty} 1\left\{\int p_{\theta}(Y_i \mid x, Z_i) \, \mathrm{d}Q_{Z_i}(x) \leq i^{-\eta}\right\} \frac{\|\hat{\theta}_{i-1} - \theta\| \int \|(x, Z_i)\| \, \mathrm{d}Q_{Z_i}(x)}{\int p_{\theta}(Y_i \mid x, Z_i) \, \mathrm{d}Q_{Z_i}(x)}
		\]
		is the sum of $\log(r_i)$ over all almost surely finitely many $i$ such that $\int p_{\theta}(Y_i \mid x, Z_i) \, \mathrm{d}Q_{Z_i}(x) \leq i^{-\eta}$. 
		Since $(X_i, Z_i)$, $i \in \mathbb{N}$, are independent and identically distributed with
		\[
		\mathbb{E}[\mathbb{E}[\|(X,Z)\||Z]^2] \leq \mathbb{E}[\|(X,Z)\|^2] < \infty,
		\]
		the law of large numbers implies 
		\[
		\frac{1}{n} \sum_{i=1}^{n} \mathbb{E}[\|(X_i,Z_i)\| \mid Z_i]^2 \leq \frac{1}{n}\sum_{i=1}^{n} \mathbb{E}[\|(X_i,Z_i)\|^2 \mid Z_i] \rightarrow \mathbb{E}[\|(X,Z)\|^2] < \infty
		\]
		almost surely as $n \rightarrow \infty$. On the other hand, $n^{2\eta}\|\hat{\theta}_{n-1}-\theta\|^2 = \mathcal{O}(n^{2\eta - 1}\log(\log(n)))$ almost surely, so that for $\eta < 1/2$, we have $n^{2\eta}\|\hat{\theta}_{n-1}-\theta\|^2 \rightarrow 0$ almost surely as $n \rightarrow \infty$. Finally, since $M$ only takes finite values, also $M/n \rightarrow 0$ for $n \rightarrow \infty$. Hence \eqref{eq:lower_bound_r} converges to $0$ almost surely.
		It follows that
		\[
		\liminf_{n\rightarrow\infty} \frac{1}{n}\left(\log(S_n^{CI}) - \log\left(\prod_{i=1}^n\frac{p_{\theta}(Y_i \mid X_i, Z_i)}{\int p_{\theta}(Y_i \mid x, Z_i)\, \mathrm{d}Q_{Z_i}(x)}\right) \right) \geq 0
		\]
		almost surely. Since 
		\[
		\frac{1}{n}\sum_{i=1}^n\log\left(\frac{p_{\theta}(Y_i \mid X_i, Z_i)}{\int p_{\theta}(Y_i \mid x, Z_i) \, \mathrm{d}Q_{Z_i}(x)}\right) \rightarrow I(X;Y\mid Z) > 0, \ n \rightarrow \infty,
		\]
		almost surely, by the law of large numbers, this proves the theorem.
	\end{proof}

	\section{Anytime-Valid \textit{E}-statistics}\label{app:loavev}
	In this section, we discuss an alternative way to define anytime-valid tests using \E-statistics and show that, in the setting of our paper, this method coincides with the method discussed in Section~\ref{sec:e_stat}. 
	In Section~\ref{sec:e_stat}, we mentioned that a sequence of conditional \E-statistics gives rise to a test martingale $(S_n(D^n))_{n\in \mathbb{N}}$, which satisfies $\mathbb{E}_P[S_\tau(D^\tau)]\leq 1$ for any stopping time $\tau$ and $P\in\mathcal{H}_0$.
	Rather than taking the latter as a consequence, \citet{koolen2022logoptimal} take this as the definition of what they call \emph{anytime-valid} \E-statistics.
	That is, they call a nonnegative process $(E_n(D^n))_{n\in \mathbb{N}}$ an anytime-valid \E-statistic if $\mathbb{E}_P[E_\tau(D^\tau)]\leq 1$ for any stopping time $\tau$ and $P\in\mathcal{H}_0$. The same object is referred to as \E-process in \citet{RamdasRLK21}, and it can be shown that the class of anytime-valid \E-statistics (or \E-processes) is strictly larger than the class of test martingales.
	A priori it is not obvious whether the GRO criterion, which maximizes the expected growth rate without referring to any particular stopping time, also yields powerful \E-statistics when specific stopping rules $\tau$ are applied.
	Therefore, \citet{koolen2022logoptimal} propose, for fixed alternative distribution $\mathcal{H}_1=\{P^*\}$ and stopping time $\tau$, to look for the anytime-valid \E-statistic that maximizes
	\begin{equation} \label{eq:e_log_tau}
	    (E_n)_{n\in \mathbb{N}}\mapsto \mathbb{E}_{P^*} [\log E_\tau(D^\tau)].
	\end{equation}
	It turns out that there are settings in which the optimal anytime-valid \E-statistic is actually equal to the GRO test martingale. 
	One of the settings in which this happens is given in their Theorem~12.
	We present a slightly rephrased version of this theorem here.
	\begin{theorem}[\citet{koolen2022logoptimal}]\label{thm:loavev}
	Assume that the data is given by an i.i.d.~stream $(D_i)_{i\in \mathbb{N}}$ and that the alternative is given by $\mathcal{H}_1=\{P^*\}$, where $P^*$ admits a density $p^*$.
	Suppose further that the GRO \E-statistic is given by the likelihood ratio $p^*/q$, where $q$ is the density of an element of $\mathcal{H}_0$. 
	Then the process $\left(p^*(D_i)/q(D_i)\right)_{i\in\mathbb{N}}$ also maximizes \eqref{eq:e_log_tau} for any stopping time $\tau$. 
	\end{theorem}
	In the proof of our Theorem~\ref{thm:GRO} (see Appendix~\ref{app:proof_GRO}), we show that the GRO \E-variable is exactly of the form described in Theorem~\ref{thm:loavev}.
	It therefore follows that the test martingale that we give in~\eqref{eq:e_information} is actually also the optimal anytime-valid \E-statistic. 
	We therefore chose to focus on the GRO property in this article.
	
\end{document}